\documentclass[12pt]{article}
\usepackage{amsmath}
\usepackage{amssymb}
\usepackage{latexsym}
\usepackage{amsthm}
\newtheorem{prop}{Proposition}

\newtheorem*{lemma*}{Lemma}
\newtheorem*{fell}{Fell's Theorem}
\newtheorem*{sv}{Stone-von Neumann Uniqueness Theorem}
\newtheorem*{gns}{Gelfand-Naimark-Segal Theorem}
\newtheorem*{fulling}{Fulling's ``Theorem''}

\theoremstyle{remark}

\theoremstyle{definition}

\newcommand{\fin}{\mathbb{F}}

\newcommand{\alg}[1]{\mathcal{#1}}
\newcommand{\norm}[1]{\| #1\|}

\newcommand{\hil}[1]{\mathcal{#1}}
\begin{document}
\begin{center}
{\Large Are Rindler Quanta Real?} \end{center} 
\begin{center}
{\Large Inequivalent Particle Concepts} \end{center} 
 \begin{center}
{\Large in} \end{center} 
 \begin{center}
{\Large Quantum Field Theory} \end{center} 
  \begin{center}
\emph{Rob Clifton and Hans Halvorson} \\
Department of Philosophy \\
 University of Pittsburgh
\end{center}
\begin{abstract}
Philosophical reflection on quantum field theory
has tended to focus on how it revises our conception
of what a particle is.  However, there has been relatively little discussion of 
the threat to the ``reality'' of particles posed by the possibility of 
inequivalent quantizations of a classical field theory, i.e., 
inequivalent representations of the algebra of observables of the field 
in terms of operators on a Hilbert space.  The threat is that each 
representation embodies its own distinctive conception of what a particle 
is, and how a ``particle'' will respond to a suitably 
operated detector. 
Our main goal is to clarify the subtle relationship between 
inequivalent representations of a field theory and their associated 
particle concepts.  We also have a particular interest 
in the Minkowski versus Rindler 
quantizations of a free Boson field, because they respectively entail two radically 
different descriptions of the particle content of the field in the 
\emph{very same} region of spacetime.  We shall defend the idea that 
these representations provide \emph{complementary descriptions} of the 
same state of the field against the claim that they embody  completely  
\emph{incommensurable} \emph{theories} of the field. \end{abstract}

\noindent --------------------------------------------------------------------------------------------------

\noindent \textbf{1} \textit{Introduction}

\noindent \textbf{2} \textit{Inequivalent 
Field Quantizations}

\textbf{2.1} \textit{The Weyl 
Algebra}

\textbf{2.2} \textit{Equivalence and Disjointness of Representations}

\textbf{2.3} \textit{Physical Equivalence of Representations}

\noindent \textbf{3} \textit{Constructing Representations}

\textbf{3.1} \textit{First Quantization (``Splitting the Frequencies'')}

\textbf{3.2} \textit{Second Quantization (Fock Space)}

\textbf{3.3} \textit{Disjointness of Minkowski and Rindler 
Representations}

\noindent \textbf{4} \textit{Minkowski versus Rindler Quanta}

\textbf{4.1} \textit{The Paradox of the Observer-Dependence of 
Particles}

\textbf{4.2} \textit{Minkowski Probabilities for Rindler Number 
Operators}

\textbf{4.3} \textit{Incommensurable or Complementary?}

\noindent \textbf{5} \textit{Conclusion}

\noindent \textit{Appendix}

\noindent --------------------------------------------------------------------------------------------------
\vspace{.1in}
\begin{quote} 
\emph{Sagredo}: Do we not see here another example of 
    that all-pervading principle of complementarity which excludes the
    simultaneous applicability of concepts to the real objects of our
    world?
    
    Is it not so that, rather than being frustrated by this 
    limitation of our conceptual grasp of the reality, we see in this 
    unification of opposites the deepest and most satisfactory result 
    of the dialectical process in our struggle for understanding? 
    
    \textit{Are Quanta Real? A Galilean Dialogue} (Jauch [1973], p. 
    48)
\end{quote}

\section{Introduction}

Philosophical reflection on quantum field theory has tended to focus
on how it revises our conception of what a particle is.  For instance,
though there is a self-adjoint operator in the theory representing the
total number of particles of a field, the standard ``Fock space''
formalism does not individuate particles from one another.  Thus,
Teller ([1995], Ch. 2) suggests that we speak of quanta that can be
``aggregated'', instead of (enumerable) particles --- which implies
that they can be distinguished and labelled.  Moreover, because the
theory \emph{does} contain a total number of quanta observable (which,
therefore, has eigenstates corresponding to different values of this
number), a field state can be a nontrivial superposition of number
eigenstates that fails to predict any particular number of quanta with
certainty.  Teller ([1995], pp. 105-6) counsels that we think of these
superpositions as not actually containing any quanta, but only
propensities to \emph{display} various numbers of quanta when the
field interacts with a ``particle detector''.

The particle concept seems so thoroughly denuded by quantum field
theory that is hard to see how it could possibly underwrite the
particulate nature of laboratory experience.  Those for whom fields
are the fundamental objects of the theory are especially aware of this
explanatory burden:

\begin{quote}
...quantum field theory is the quantum theory of a field, not a theory of
``particles''. However, when we consider the manner in which a quantum field
interacts with other systems to which it is coupled, an interpretation of
the states in [Fock space] in terms of ``particles'' naturally arises. It
is, of course, essential that this be the case if quantum field
theory is to describe observed phenomena, since ``particle-like'' behaviour
is commonly observed (Wald [1994], pp. 46-7).
\end{quote}

\noindent These remarks occur in the context of Wald's discussion of 
yet another threat to the ``reality'' of quanta.  

The threat arises from the possibility of inequivalent representations
of the algebra of observables of a field in terms of operators on a
Hilbert space.  Inequivalent representations are required in a variety
of situations; for example, interacting field theories in which the
scattering matrix does not exist (``Haag's theorem''), free fields
whose dynamics cannot be unitarily implemented (Arageorgis \emph{et
  al} [2000]), and states in quantum statistical mechanics
corresponding to different temperatures (Emch [1972]).  The catch is
that each representation carries with it a distinct notion of
``particle''.  Our main goal in this paper is to clarify the subtle
relationship between inequivalent representations of a field theory
and their associated particle concepts.

Most of our discussion shall apply to any case in which inequivalent
representations of a field are available.  However, we have a
particular interest in the case of the Minkowski versus Rindler
representations of a free Boson field.  What makes this case
intriguing is that it involves two radically different descriptions of
the particle content of the field in the \emph{very same} spacetime
region.  The questions we aim to answer are:
\begin{itemize}
\item Are the Minkowski and Rindler descriptions nevertheless, in some sense, \emph{physically}
equivalent?  
\item Or, are they incompatible, even theoretically \emph{incommensurable}?  
\item Can they be thought of
as \emph{complementary} descriptions in the same way that the concepts of position
and momentum are?  
\item Or, can at most one description, the ``inertial'' story in terms
  Minkowski quanta, be the \emph{correct} one?
\end{itemize} 

Few discussions of Minkowski versus Rindler quanta broaching these
questions can be found in the philosophical literature, and what
discussion there is has not been sufficiently grounded in a rigorous
mathematical treatment to deliver cogent answers (as we shall see).
We do not intend to survey the vast physics literature about Minkowski
versus Rindler quanta, nor all physical aspects of the problem.  Yet a
proper appreciation of what is at stake, and which answers to the
above questions are sustainable, requires that we lay out the basics
of the relevant formalism. We have strived for a self-contained
treatment, in the hopes of opening up the discussion to philosophers
of physics already familiar with elementary non-relativistic quantum
theory. (We are inclined to agree with Torretti's recent diagnosis
that most philosophers of physics tend to neglect quantum field theory
because they are ``sickened by untidy math'' ([1999], p.  397).)
 
We begin in section \textbf{2} with a general introduction to the
problem of quantizing a classical field theory.  This is followed by a
detailed discussion of the conceptual relationship between
inequivalent representations in which we reach conclusions at variance
with some of the extant literature.  In section \textbf{3}, we explain
how the state of motion of an observer is taken into account when
constructing a Fock space representation of a field, and how the
Minkowski and Rindler constructions give rise to inequivalent
representations.  Finally, in section \textbf{4}, we examine the
subtle relationship between the different particle concepts implied by
these representations.  In particular, we defend the idea that they
supply \emph{complementary} descriptions of the same field against the
claim that they embody different, incommensurable \emph{theories}.
 
A certain number of mathematical results play an important role in our
exposition and in our philosophical arguments.  The results are stated
in the main text as propositions, and the proofs of those that cannot
be found in the literature are included in an appendix.

\section{Inequivalent Field Quantizations}
In section \textbf{2.1} we discuss the Weyl algebra, which in the case
of infinitely many degrees of freedom circumscribes the basic
kinematical structure of a free Boson field.  After introducing in
section \textbf{2.2} some important concepts concerning
representations of the Weyl algebra in terms of operators on Hilbert
space, we shall be in a position to draw firm conclusions about the
conceptual relation between inequivalent representations in section
\textbf{2.3}.
 
\subsection{The Weyl Algebra}
Consider how one constructs the quantum-mechanical analogue of a classical 
system with a finite number of degrees of freedom, described by a $2n$-dimensional 
phase space $S$.  
Each point of $S$ is determined by a pair of vectors 
$\vec{a},\vec{b}\in\mathbb{R}^{n}$ whose components $\{a_{j}\}$ and $\{b_{j}\}$  encode all the position and 
momentum components of the system
\begin{equation}
x(\vec{a})=\sum_{j=1}^{n}a_{j}x_{j},\ p(\vec{b})=\sum_{j=1}^{n}b_{j}p_{j}.
\end{equation}  
To quantize the system, we impose the \emph{canonical 
commutation relations} (CCRs)
\begin{equation} \label{eq:glob}
[x(\vec{a}),x(\vec{a}')]=[p(\vec{b}),p(\vec{b}')]=0,\ 
[x(\vec{a}),p(\vec{b})]=i(\vec{a}\cdot\vec{b})I,
\end{equation}
and, then, seek a representation of these relations in terms of 
operators on a Hilbert space $\hil{H}$.  In the standard \emph{Schr\"{o}dinger 
representation}, $\hil{H}$ is the space of square-integrable wavefunctions 
$L^{2}(\mathbb{R}^{n})$, $x(\vec{a})$ becomes the operator that 
multiplies a wavefunction $\Psi(x_{1},\ldots,x_{n})$ 
by $\sum_{j=1}^{n}a_{j}x_{j}$, and $p(\vec{b})$ is the partial 
differential operator $-i\sum_{j=1}^{n}b_{j}\frac{\partial}{\partial 
x_{j}}$.  

Note the action of $x(\vec{a})$ is not defined on an element $\Psi\in
L^{2}(\mathbb{R}^{n})$ unless $x(\vec{a}) \Psi$ is again
square-integrable, and $p(\vec{b})$ is not defined on $\Psi$ unless it
is suitably differentiable.  This is not simply a peculiarity of the
Schr\"{o}dinger representation.  Regardless of the Hilbert space on
which they act, two self-adjoint operators whose commutator is a
nonzero scalar multiple of the identity, as in (\ref{eq:glob}), cannot
both be everywhere defined (Kadison \& Ringrose (henceforth, KR)
[1997], Remark 3.2.9).  To avoid the technical inconvenience of
dealing with domains of definition, it is standard to reformulate the
representation problem in terms of unitary operators which are
bounded.
 
Introducing the two $n$-parameter families of unitary operators
\begin{equation} \label{eq:blob}
U(\vec{a}):=e^{ix(\vec{a})},\ V(\vec{b}):=e^{ip(\vec{b})},\qquad 
\vec{a},\vec{b}\in\mathbb{R}^{n},
\end{equation}
 it can be shown, at least formally, that the
  CCRs are equivalent to
 \begin{equation}\label{eq:wfCCR1}
 U(\vec{a})U(\vec{a}')=U(\vec{a}+\vec{a}'),\  V(\vec{b})V(\vec{b}')=V(\vec{b}
 +\vec{b}'), 
 \end{equation}
 \begin{equation}\label{eq:wfCCR2}
 U(\vec{a})V(\vec{b})=
 e^{i(\vec{a}\cdot\vec{b})}V(\vec{b})U(\vec{a}),
 \end{equation}
 called the \emph{Weyl form} of the CCRs.  This equivalence holds
 rigorously in the Schr\"{o}dinger representation, however there are
 ``irregular'' representations in which it fails (see Segal [1967],
 Sec. 1; Summers [1998], Sec. 1).  Thus, one reconstrues the goal as
 that of finding a representation of the Weyl form of the CCRs in
 terms of two concrete families of unitary operators
 $\{U(\vec{a}),V(\vec{b}): \vec{a},\vec{b}\in\mathbb{R}^{n}\}$ acting
 on a Hilbert space $\hil{H}$ that \emph{can} be related, via
 (\ref{eq:blob}), to canonical position and momentum operators on
 $\hil{H}$ satisfying the CCRs.  We shall return to this latter
 ``regularity'' requirement later in this section.
  
 Though the position and momentum degrees of freedom 
 have so far been treated on a different footing, we can simplify things 
 further by introducing the composite \emph{Weyl 
 operators} 
 \begin{equation} 
 W(\vec{a},\vec{b}):=e^{i(\vec{a}\cdot\vec{b})/2}V(\vec{b})U(\vec{a}), 
 \qquad \vec{a},\vec{b}\in \mathbb{R}. \end{equation}
 Combining this definition with Eqns. (\ref{eq:wfCCR1}) and 
 (\ref{eq:wfCCR2})  yields the 
multiplication rule
\begin{equation} \label{eq:m} W(\vec{a},\vec{b})W(\vec{a}',\vec{b}')=
e^{-i\sigma((\vec{a},\vec{b}),(\vec{a}',\vec{b}'))/2}W(\vec{a}+\vec{a}',\vec{b}+
\vec{b}'),\end{equation} 
where 
\begin{equation}
\sigma((\vec{a},\vec{b}),(\vec{a}',\vec{b}')):=(\vec{a}'\cdot
\vec{b})-(\vec{a}\cdot
\vec{b}').  
\end{equation}
Observe that $\sigma(\cdot,\cdot)$ is an anti-symmetric, bilinear 
form on $S$, called a \emph{symplectic} form.  (Note, also, that $\sigma$ 
is nondegenerate; i.e., if for any $f\in S$, $\sigma(f,g)=0$ for all $g\in S$, then 
 $f=0$.) We set 
\begin{equation} \label{eq:m'} 
W(\vec{a},\vec{b})^{*}:=e^{-i(\vec{a}\cdot\vec{b})/2}U(-\vec{a})V(-\vec{b})=
W(-\vec{a},-\vec{b}).\end{equation}  Clearly, then, 
any representation of the Weyl operators $W(\vec{a},\vec{b})$ on a
Hilbert space $\hil{H}$ gives rise to a representation of the Weyl 
form of the CCRs,
and vice-versa.     
 
Now, more generally, we allow our classical phase space $S$ to be any
infinite-dimensional vector space, possibly constructed out of
solutions to some relativistic wave equation. We assume $S$ comes
equipped with a (nondegenerate) symplectic form $\sigma$, and we say
that a family $\{ W_{\pi}(f):f\in S\}$ of unitary operators acting on
some Hilbert space $\hil{H}_{\pi}$ satisfies \emph{the Weyl relations}
just in case (cf. (\ref{eq:m}), (\ref{eq:m'}))
\begin{equation} \label{eq:herro} W_{\pi}(f)W_{\pi}(g)=e^{-i\sigma (f,g)/2}W_{\pi}(f+g) , 
\qquad f,g\in
  S,\end{equation} 
  \begin{equation}
  W_{\pi}(f)^{*}=W_{\pi}(-f), 
\qquad f\in
  S.
  \end{equation}
We may go on to form arbitrary linear combinations of the Weyl
operators, and thus obtain (at least some of) the self-adjoint
operators that will serve as observables of the system.

Let $\alg{F}$ be a family of bounded operators on $\hil{H}_{\pi}$.  We
say that a bounded operator $A$ on $\hil{H}_{\pi}$ may be 
\emph{uniformly}
  approximated by operators in $\alg{F}$ just in case for every
$\epsilon >0$, there is an operator $\tilde{A}\in \alg{F}$ such that
\begin{equation}
\norm{(A-\tilde{A})x}<\epsilon ,\ \mbox{for all unit vectors}\ 
x\in \hil{H}_{\pi}. \label{uniform} \end{equation} 
  Let $\alg{W}_{\pi}$ 
denote the set of all bounded operators on $\hil{H}_{\pi}$ that can be uniformly
approximated by elements in $\alg{F}$, where $\alg{F}$ is the set of
linear combinations of Weyl operators $W_{\pi}(f)$ acting on
$\hil{H}_{\pi}$.  $\alg{W}_{\pi}$ is called the $C^{*}$-\emph{algebra}
generated by the Weyl operators $\{ W_{\pi}(f)\}$.  In particular, 
$\alg{W}_{\pi}$ is a subalgebra of the algebra 
$\mathbf{B}(\hil{H}_{\pi})$ of \emph{all} bounded operators on
$\hil{H}_{\pi}$ that is itself uniformly closed and closed under taking adjoints
$A\mapsto A^{*}$.        

Suppose, now, that $\{ W_{\pi}(f)\}$ and $\{ W_{\phi}(f)\}$ are systems
of Weyl operators acting, respectively, on Hilbert spaces
$\hil{H}_{\pi},\hil{H}_{\phi}$.  Let $\alg{W}_{\pi},\alg{W}_{\phi}$
denote the corresponding $C^{*}$-algebras.  A bijective mapping
$\alpha :\alg{W}_{\pi}\mapsto \alg{W}_{\phi}$ is called a
$*$-\emph{isomorphism} just in case $\alpha$ is linear,
multiplicative, and commutes with the adjoint operation.  We then have
the following uniqueness result for the $C^{*}$-algebra generated by
Weyl operators (see Bratteli \& Robinson (henceforth, BR)
[1996], Thm. 5.2.8).

\begin{prop} There is a $*$-isomorphism $\alpha$ from $\alg{W}_{\pi}$ onto
  $\alg{W}_{\phi}$ such that $\alpha (W_{\pi}(f))=W_{\phi}(f)$ for all
  $f\in S$.  \end{prop}

\noindent This establishes that the $C^{*}$-algebra generated in any 
representation by Weyl operators satisfying the Weyl relations is, in 
fact, 
representation-independent.  We shall denote this abstract algebra, 
called the \emph{Weyl
  algebra over} $(S,\sigma)$, by $\alg{W}[S,\sigma]$ 
  (and, when no confusion
can result, simply say ``Weyl algebra'' and write $\alg{W}$ for 
$\alg{W}[S,\sigma ]$).  
So our problem boils down to choosing a 
\emph{representation} $(\pi ,\hil{H}_{\pi})$ of the Weyl algebra, 
given by a mapping $\pi:\alg{W}[S,\sigma ]\mapsto\mathbf{B}(\hil{H}_{\pi})$ 
preserving all algebraic relations. Note, also, 
that 
since the image $\pi(\alg{W})$ will always be an isomorphic copy of 
$\alg{W}$, $\pi$ will always be one-to-one, and hence provide a  
\emph{faithful} representation of $\alg{W}$.   

With the representation-independent character of the Weyl algebra
$\alg{W}$, why should we care any longer to choose a representation?  
After all, there is no technical obstacle to 
proceeding abstractly.  We can take the self-adjoint elements of $\alg{W}$ 
to be the quantum-mechanical observables of our system.  
A linear functional $\omega$ on $\alg{W}$ is
called a \emph{state} just in case $\omega$ is positive (i.e., $\omega
(A^{*}A)\geq 0$) and normalized (i.e., $\omega (I)=1$).  As usual, a 
state $\omega$ is taken to be \emph{pure} (and \emph{mixed} 
otherwise) just in case it is not a 
nontrivial convex combination of other states of $\alg{W}$.  The dynamics
of the system can be represented by a one-parameter group $\alpha _{t}$ of
automorphisms of $\alg{W}$ (i.e., each $\alpha _{t}$ is just a map
of $\alg{W}$ onto itself that preserves all algebraic relations).  
Hence, if we have some 
initial state $\omega_{0}$, the final state will be given by 
$\omega_{t}=\omega_{0}\circ \alpha _{t}$.   We 
can even supply definitions for the probability in the state 
$\omega_{t}$ 
that a self-adjoint element $A\in\alg{W}$ takes a value lying in some 
Borel
subset of its spectrum (Wald [1994], pp. 79-80), and for 
transition probabilities between, and superpositions of, pure states 
of 
$\alg{W}$ (Roberts \& Roepstorff [1969]).  
At no stage, it seems, need we ever 
introduce a Hilbert space as 
an essential element of the formalism. 
In fact, 
Haag and Kastler ([1964], p. 852) and Robinson ([1966], p. 
488) maintain that the choice of a 
representation is largely a matter of analytical convenience without 
physical implications.

Nonetheless, the abstract Weyl algebra does not contain
unbounded operators, many of which are naturally taken as
corresponding to important physical quantities.  For instance, 
the total energy of the system, the canonically conjugate position and momentum 
observables --- which in field theory play the role of the local field 
observables --- and the total
number of particles.  Also, we shall see later that not even any 
\emph{bounded} function of the total number of particles (apart from 
zero and the identity) 
lies in the Weyl algebra.  Surprisingly, Irving Segal (founder of the 
mathematically rigorous approach 
to quantum field theory) has written that this: 
\begin{quote}
\ldots has the simple if quite
rough and somewhat oversimplified interpretation that the total number of
``bare'' particles is devoid of physical meaning (Segal [1963], p. 
56; see also his [1959], p. 12). 
\end{quote}
\noindent We shall return to this issue of 
physical meaning shortly.  First, let us see how a 
representation can be used to 
expand the observables of a system beyond the abstract Weyl algebra. 

Let $\alg{F}$ be a family of bounded operators acting on a 
representation space
$\hil{H}_{\pi}$.  We say that a bounded operator $A$ on
$\hil{H}_{\pi}$ can be \emph{weakly} approximated by elements of
$\alg{F}$ just in case for any vector $x\in \hil{H}$, and any
$\epsilon >0$, there is some $\tilde{A}\in \alg{F}$ such that
\begin{equation} \left| \langle x,Ax\rangle -\langle x,\tilde{A}x\rangle
  \right| < \epsilon .
\end{equation}  (Note the important quantifier change between the
definitions of uniform and weak approximation, and that weak 
approximation has no abstract representation-independent counterpart.)  
Consider the family $\pi(\alg{W})^{-}$ of bounded operators that can be weakly 
approximated by elements of $\pi(\alg{W})$, i.e., $\pi(\alg{W})^{-}$ 
is 
the weak closure of $\pi(\alg{W})$.   By von Neumann's  
double commutant theorem, 
$\pi(\alg{W})^{-}=\pi(\alg{W})''$, where the prime operation on a family of 
operators (here applied twice) denotes the set of all bounded 
operators on $\hil{H}_{\pi}$ commuting with that family.    
$\pi(\alg{W})''$ is called the \emph{von Neumann algebra} generated 
by $\pi(\alg{W})$.  Clearly
$\pi(\alg{W})\subseteq \pi(\alg{W})''$, however we can hardly
expect that $\pi(\alg{W})=\pi(\alg{W})''$ when $\hil{H}_{\pi}$ is 
infinite-dimensional (which it \emph{must} be, since there is no 
finite-dimensional representation of the Weyl algebra for even a 
single degree of freedom).  Nor should we generally expect that 
$\pi(\alg{W})''=\mathbf{B}(\hil{H}_{\pi})$, though this does hold in ``irreducible'' 
representations, as we explain in the next subsection.

We may now expand our observables to include all self-adjoint 
operators in $\pi(\alg{W})''$.  And, although 
$\pi(\alg{W})''$ still contains only bounded operators, it
is easy to associate (potentially physically significant) 
unbounded observables with this algebra as well.  We 
say that a (possibly unbounded) self-adjoint operator $A$ on
$\hil{H}_{\pi}$ is \emph{affiliated} with $\pi(\alg{W})''$ just in case
all $A$'s spectral projections lie in $\pi(\alg{W})''$.  Of course, we 
could have adopted the same definition for self-adjoint operators 
``affiliated to'' $\pi(\alg{W})$  itself, but $C^{*}$-algebras do not 
generally contain nontrivial projections (or, if they do, will not 
generally contain even
the spectral projections of their self-adjoint 
members).

As an example, suppose 
we now demand to have a (so-called) \emph{regular} 
representation $\pi$, in which the mappings 
$t\in\mathbb{R}\mapsto \pi(W(tf))$, for all $f\in S$, are all weakly continuous. 
Then Stone's theorem will guarantee
 the existence 
of unbounded self-adjoint operators $\{\Phi(f):f\in S\}$ on $\hil{H}_{\pi}$ satisfying 
$\pi(W(tf))=e^{i\Phi(f)t}$, and it can be shown that all these 
operators are affiliated to $\pi(\alg{W})''$ (KR [1997], Ex. 5.7.53(ii)). 
In this way, we can recover 
as observables our original canonically conjugate positions and 
momenta (cf. Eqn. (\ref{eq:blob})), which the Weyl relations 
ensure will satisfy the original 
unbounded form of the CCRs.  

It is important to recognize, however, that by enlarging the set of
observables to include those affiliated to $\pi(\alg{W})''$, we have
now left ourselves open to arbitrariness. In contrast to Proposition
1, we now have
\begin{prop} There are (even regular) representations $\pi ,\phi$ 
of $\alg{W}[S,\sigma]$ for which there \emph{is no} $*$-isomorphism 
$\alpha$ from $\pi(\alg{W})''$ onto
  $\phi(\alg{W})''$ such that $\alpha (\pi(W(f)))=\phi(W(f))$ for all
  $f\in S$. \end{prop}
  \noindent This occurs when the representations are ``disjoint'', 
  which we discuss in the next subsection. 
  
  Proposition 2 is what motivates Segal to argue that observables
  affiliated to the weak closure $\pi(\alg{W})''$ in a representation
  of the Weyl algebra are ``somewhat unphysical'' and ``have only
  analytical significance'' ([1963], pp. 11--14,
  134).\footnote{Actually, Segal consistently finds it convenient to
    work with a strictly larger algebra than our (minimal) Weyl
    algebra, sometimes called the \emph{mode finite} or \emph{tame}
    Weyl algebra. However, both Proposition 1 (see Baez \emph{et al}
    [1992], Thm. 5.1) and Proposition 2 continue to hold for the tame
    Weyl algebra (also cf.  Segal [1967], pp. 128-9).}  Segal is
  explicit that by ``physical'' he means ``empirically measurable in
  principle'' ([1963], p.  11). We should not be confused by the fact
  that he often calls observables that fail this test ``conceptual''
  (suggesting they are more than mere analytical crutches).  For in
  Baez \emph{et al} ([1992], p. 145), Segal gives as an example the
  bounded self-adjoint operator $\cos p +(1+x^{2})^{-1}$ on
  $L^{2}(\mathbb{R})$ ``for which no known `Gedanken experiment' will
  actually directly determine the spectrum, and so [it] represents an
  observable in a purely conceptual sense''.  Thus, the most obvious
  reading of Segal's position is that he subscribes to an
  operationalist view about the physical significance of theoretical
  quantities.  Indeed, since good reasons \emph{can} be given for the
  impossibility of exact (``sharp'') measurements of all the
  observables in a von Neumann algebra generated by a $C^{*}$-algebra
  (see Wald [1994], Halvorson [2000a]), operationalism explains
  Segal's dismissal of the physical (as opposed to analytical)
  significance of observables not in the Weyl algebra \emph{per se}.
  (Also, it is worth recalling that Bridgman himself was similarly
  unphased by having to relegate much of the mathematical structure of
  a physical theory to ``a ghostly domain with no physical relevance''
  ([1936], p. 116).)
  
  Of course, insofar as operationalism is philosophically 
  defensible at all, it does not compell assent.  And, in this instance, 
  Segal's operationalism has not dissuaded
  others from taking 
  the more liberal view apparently advocated by Wald: 
  \begin{quote}
\ldots one should not view [the Weyl algebra] as encompassing \textit{all}
observables of the theory; rather, one should view [it] as encompassing a
``minimal'' collection of observables, which is sufficiently large to enable
the theory to be formulated. One may later wish to enlarge [the algebra]
and/or further restrict the notion of ``state'' in order to accommodate the
existence of additional observables ([1994], p. 75).
\end{quote}
\noindent The conservative and liberal views 
entail quite different commitments about the physical equivalence 
of representations --- or so we shall argue.  

\subsection{Equivalence and Disjointness of Representations}
It is essential that precise mathematical definitions of 
equivalence be clearly distinguished from the, often dubious, arguments that 
have been offered for their conceptual significance.  We confine this 
section to discussing the definitions. 

Since our ultimate goal is to discuss the 
Minkowski and Rindler quantizations of the Weyl algebra, we only need 
to consider the case where one of the two representations at 
issue, say $\pi$, is ``irreducible'' and the other, $\phi$, is ``factorial''. 
A representation $\pi$ of $\alg{W}$ is called \emph{irreducible}
just in case no non-trivial subspace of the Hilbert space
$\hil{H}_{\pi}$ is invariant under the action of all operators
in $\pi(\alg{W})$.  It is not difficult to see that this is equivalent to 
$\pi(\alg{W})''=\mathbf{B}(\hil{H}_{\pi})$ (using the fact that 
an invariant subspace will exist just in case 
the projection onto it commutes with all of $\pi(\alg{W})$). 
A representation $\phi$ of $\alg{W}$ is called \emph{factorial}
whenever the von Neumann algebra $\phi(\alg{W})''$ is a \emph{factor}, 
i.e., it has trivial centre (the only operators in $\phi(\alg{W})''$ that 
commute with all other operators in that set are multiples of the 
identity).  Since $\mathbf{B}(\hil{H}_{\pi})$ is a factor, it is 
clear that $\pi$'s irreducibility entails its factoriality.  
Thus, the Schr\"{o}dinger representation of the Weyl algebra is both 
irreducible and factorial.   

The strongest form of equivalence between 
representations is unitary equivalence: $\phi$ and $\pi$ are said to
be \emph{unitarily equivalent} just in case there is a unitary
operator $U$ mapping $\hil{H}_{\phi}$ isometrically onto
$\hil{H}_{\pi}$, and such that 
\begin{equation}
U\phi(A)U^{-1}=\pi(A)\qquad \forall A\in
\alg{W}.
\end{equation}  There are two other weaker definitions of equivalence.
 
Given a family $\pi _{i}$ of irreducible representations of the Weyl
algebra on Hilbert spaces $\hil{H}_{i}$, we can construct another
(reducible) representation $\phi$ of the Weyl algebra on the direct
sum Hilbert space $\sum \oplus \hil{H}_{i}$, by setting
\begin{equation} \phi (A)=\sum _{i}\oplus \,\pi _{i}(A) ,\qquad A\in
\alg{W}. \end{equation}
If each representation $(\pi _{i} ,\hil{H}_{i})$ is unitarily
equivalent to some representation $(\pi ,\hil{H})$, we say that $\phi
= \sum \oplus \pi _{i}$ is a \emph{multiple} of the representation $\pi$.
Furthermore, we say that two
representations of the Weyl algebra, $\phi$ (factorial) and $\pi$ 
(irreducible), 
are \emph{quasi-equivalent} just in case $\phi$ is a multiple of
$\pi$.  It should be obvious from this characterization that quasi-equivalence 
weakens unitary equivalence.

Another way to see this is to use the fact (KR [1997], Def. 10.3.1, 
  Cor. 10.3.4) that quasi-equivalence of $\phi$ and $\pi$ is 
equivalent to the existence of a $*$-isomorphism 
$\alpha$ from $\phi(\alg{W})''$ onto
  $\pi(\alg{W})''$ such that $\alpha (\phi(A))=\pi(A)$ for all
  $A\in \alg{W}$.  Unitary equivalence is then just the special case 
  where the $*$-isomorphism $\alpha$ can be implemented by a unitary operator.  

If $\phi$ is not even quasi-equivalent to $\pi$, then we say that
$\phi$ and $\pi$ are \emph{disjoint} representations of
$\alg{W}$.\footnote{In general, disjointness is not defined as the
  negation of quasi-equivalence, but by the more cumbersome
  formulation: Two representations $\pi ,\phi$ are disjoint just in
  case $\pi$ has no ``subrepresentation'' quasi-equivalent to $\phi$,
  and $\phi$ has no subrepresentation quasi-equivalent to $\pi$.  
  Since we
  are only interested, however, in the special case where $\pi$ is
  irreducible (and hence has no non-trivial subrepresentations) and
  $\phi$ is ``factorial'' (and hence is quasi-equivalent to each of
  its subrepresentations), the cumbersome formulation reduces to our 
  definition.} Note, then, that if both $\pi$ and $\phi$
are irreducible, they are either unitarily equivalent or
disjoint.  

We can now state the following pivotal result (von Neumann [1931]).
\begin{sv} When $S$ is finite-dimensional, every regular 
representation of the Weyl algebra $\alg{W}[S,\sigma]$ is 
quasi-equivalent to the Schr\"{o}dinger representation. 
\end{sv}
\noindent This theorem is usually interpreted
as saying that there is a unique quantum theory corresponding
to a classical theory with finitely-many degrees of freedom.     
The theorem \emph{fails} in field theory --- where $S$ is 
infinite-dimensional --- opening the door to disjoint 
representations and Proposition 2.  

There is another way to think of the relations between representations,
in terms of states.  Recall the abstract definition of a 
state of a $C^{*}$-algebra, as simply 
a positived normalized linear functional on the algebra. 
Since, in any representation $\pi$, $\pi (\alg{W})$ is just a 
faithful copy of 
$\alg{W}$, $\pi$ induces a one-to-one correspondence between the abstract
states of $\alg{W}$ and the abstract states of $\pi (\alg{W})$.
Note now
that \emph{some} of the abstract states on $\pi (\alg{W})$ are the
garden-variety density operator states that we are familiar with from
elementary quantum mechanics.  In particular, define $\omega _{D}$ on $\pi (\alg{W})$
by setting
\begin{equation} \omega _{D}(A):=\mathrm{Tr}(DA) ,\qquad A\in \pi
  (\alg{W}). \label{normal}
\end{equation}
\emph{In general, however, there will be abstract states of $\pi
  (\alg{W})$ that are not given by density operators via
  Eqn.~(\ref{normal}).}\footnote{\label{count} Gleason's theorem does
  not rule out these states because it is not part of the definition
  of an abstract state that it be countably additive over mutually
  orthogonal projections.  Indeed, such additivity does not even make
  sense abstractly, because an infinite sum of orthogonal projections
  can never converge uniformly, only weakly (in a representation).} We
say then that an abstract state $\omega$ of $\pi (\alg{W})$ is
\emph{normal} just in case it is given (via Eqn.~(\ref{normal})) by
some density operator $D$ on $\hil{H}_{\pi}$.  We let
$\mathfrak{F}(\pi )$ denote the subset of the abstract state space of
$\alg{W}$ consisting of those states that correspond to normal states
in the representation $\pi$, and we call $\mathfrak{F}(\pi )$ the
\emph{folium} of the representation $\pi$.  That is, $\omega \in
\mathfrak{F}(\pi )$ just in case there is a density operator $D$ on
$\hil{H}_{\pi}$ such that \begin{equation} \omega (A)=\mathrm{Tr}(D\pi
  (A)) ,\qquad A\in \alg{W} .\end{equation} We then have the following
equivalences (KR [1997], Prop. 10.3.13):
\begin{eqnarray} \nonumber
\pi\ \mbox{and}\ \phi\ \mbox{are quasi-equivalent}  &
 \Longleftrightarrow
& \mathfrak{F}(\pi )=\mathfrak{F}(\phi),  \\ \nonumber
\pi\ \mbox{and}\ \phi\ \mbox{are disjoint} & \Longleftrightarrow & 
\mathfrak{F}(\pi
  )\cap \mathfrak{F}(\phi )=\emptyset. \end{eqnarray} 
\noindent In other words, $\pi$ and $\phi$ are quasi-equivalent just in case
they share the same normal states.  And $\pi$ and $\phi$ are disjoint
just in case they have \emph{no} normal states in common.  

In fact, if $\pi$
is disjoint from $\phi$, then all normal states in the representation
$\pi$ are ``orthogonal'' to all normal states in the representation
$\phi$.  We may think of this situation intuitively as follows.  Define
a third representation $\psi$ of $\alg{W}$ on $\hil{H}_{\pi}\oplus
\hil{H}_{\phi}$ by setting
\begin{equation} \psi (A)=\pi (A)\oplus
  \phi (A) ,\qquad A\in \alg{W} .\end{equation} 
Then, every normal state of the representation $\pi$ is orthogonal to every normal state
of the representation $\phi$.\footnote{This intuitive picture may be
justified by making use of the ``universal representation'' of
$\alg{W}$ (KR [1997], Thm. 10.3.5).}  This makes sense of the oft-repeated 
phrase 
(see, e.g., Gerlach [1989]) 
that ``The Rindler vacuum is orthogonal to all states in the 
Minkowski vacuum representation''. 

While not every abstract state of $\alg{W}$ will be in the folium of a 
given representation, there is always \emph{some} representation of $\alg{W}$ in which 
the state
\emph{is} normal,  as a consequence of the following (see KR [1997],
Thms. 4.5.2 and 10.2.3).
\begin{gns} Any abstract state $\omega$ of a $C^{*}$-algebra
 $\alg{A}$ gives rise to
a unique (up to unitary equivalence) representation $(\pi
_{\omega},\hil{H}_{\omega})$ of $\alg{A}$ and vector $\Omega
_{\omega}\in \hil{H}_{\omega}$ such that
\begin{equation} \omega (A)=\langle \Omega _{\omega},\pi
  _{\omega}(A)\Omega _{\omega} \rangle ,\qquad A\in \alg{A} ,\end{equation}
and such that the set $\{ \pi _{\omega}(A)\Omega _{\omega}:A\in
\alg{A}\}$ is dense in $\hil{H}_{\omega}$.     Moreover, $\pi
_{\omega}$ is irreducible just in case $\omega$ is pure.  
\end{gns}
\noindent The triple $(\pi
_{\omega},\hil{H}_{\omega},\Omega _{\omega})$ is
called the \emph{GNS representation} of $\alg{A}$ induced by the state
$\omega$, and  $\Omega _{\omega}$ is called a \emph{cyclic} vector for 
the representation.   We shall see in the next main section how the Minkowski and Rindler 
vacuums induce disjoint GNS representations of the Weyl algebra.   

There is a third notion of equivalence of representations, still weaker than 
quasi-equivalence.  Let $\pi$ be a representation of $\alg{W}$, and let
$\mathfrak{F}(\pi)$ be the folium of $\pi$.  We say that an abstract
state $\omega$ of $\alg{W}$ can be \emph{weak* approximated} by states
in $\mathfrak{F}(\pi)$ just in case for each $\epsilon >0$, and for
each finite collection $\{ A_{i}:i=1,\dots ,n\}$ of operators in
$\alg{W}$, there is a state $\omega '\in \mathfrak{F}(\pi )$ such that
\begin{equation} \left| \omega (A_{i})-\omega '(A_{i}) \right|
  <\epsilon , \qquad i\in [1,n].\end{equation}
Two representations $\pi ,\phi$ are then said to be \emph{weakly equivalent} just
in case all states in $\mathfrak{F}(\pi )$ may be weak* approximated by
states in $\mathfrak{F}(\phi )$ and vice-versa.  We then have the 
following fundamental result (Fell [1960]).  
\begin{fell}  Let $\pi$ be a faithful representation of a $C^{*}$-algebra
 $\alg{A}$.  Then,
  every abstract state of $\alg{A}$ may be weak* approximated by
  states in $\mathfrak{F}(\pi )$.  \end{fell}
\noindent In particular, then, it follows that \emph{all} representations of
$\alg{W}$ are weakly equivalent.

In summary, we have the following implications for any two
representations $\pi ,\phi$: 
\begin{equation*} \mbox{Unitarily equivalent}\ 
\Longrightarrow\ \mbox{Quasi-equivalent}\ \Longrightarrow\ 
\mbox{Weakly equivalent}. \end{equation*} 
    If $\pi$ and $\phi$ are both
irreducible, then the first arrow is reversible. 

\subsection{Physical Equivalence of Representations}
Do disjoint representations yield \emph{physically} inequivalent
theories?  It depends on what one takes to be the physical content of
a theory, and what one means by ``equivalent theories'' --- subjects
about which philosophers of science have had plenty to say.

Recall that Reichenbach [1938] deemed two theories ``the same'' just
in case they are empirically equivalent, i.e., they are confirmed
equally under all possible evidence.  Obviously this criterion, were
we to adopt it here, would beg the question against those who (while
agreeing that, strictly speaking, only self-adjoint elements of the
Weyl algebra can actually be measured) invest physical importance to
``global'' quantities only definable in a representation, like the
total number of particles.

A stronger notion of equivalence, due originally to Glymour [1971]
(who proposed it only as a \emph{necessary} condition), is that two
theories are equivalent only if they are ``intertranslatable''.  This
is often cashed out in logical terms as the possibility of defining
the primitives of one theory in terms of those of the other so that
the theorems of the first appear as logical consequences of those of
the second, and vice-versa.  Prima facie, this criterion is ill-suited
to the present context, because the different ``theories'' are not
presented to us as syntactic structures or formalized logical systems,
but rather two competing algebras of observables whose states
represent physical predictions.  In addition, intertranslatability
\emph{per se} has nothing to say about what portions of the
mathematical formalism of the two physical theories being compared
ought to be intertranslatable, and what should be regarded as
``surplus mathematical structure'' not required to be part of the
translation.

Nevertheless, we believe the intertranslatability thesis can be
naturally expressed in the present context and rendered neutral as
between the conservative and liberal approaches to physical
observables discussed earlier.  Think of the Weyl operators
$\{\phi(W(f)):f\in S\}$ and $\{\pi(W(f)):f\in S\}$ as the primitives
of our two ``theories'', in analogy with the way the natural numbers
can be regarded as the primitives of a ``theory'' of real numbers.
Just as we may define rational numbers as ratios of natural numbers,
and then construct real numbers as the limits of Cauchy sequences of
rationals, we construct the Weyl algebras $\phi(\alg{W})$ and
$\pi(\alg{W})$ by taking linear combinations of the Weyl operators and
then closing in the uniform topology.  We then close in the weak
topology of the two representations to obtain the von Neumann algebras
$\phi(\alg{W})''$ and $\pi(\alg{W})''$.  Whether the observables
affiliated with this second closure have physical significance is up
for grabs, as is whether we should be conservative and take only
normal states in the given representation to be physical, or be more
liberal and admit a broader class of algebraic states.  The analogue
of the ``theorems'' of the theory are then statements about the
expectation values dictated by the physical states for the
self-adjoint elements in the physically relevant algebra of the
theory.

We therefore propose the following formal rendering of Glymour's
inter-translatability thesis adapted to the present context.
Representations $\phi$ and $\pi$ are \emph{physically equivalent} only
if there exists a bijective mapping $\alpha$ from the physical
observables of the representation $\phi$ to the physical observables
of the representation $\pi$, and another bijective mapping $\beta$
from the physical states of the representation $\phi$ to the physical
states of the representation $\pi$, such that \begin{eqnarray}
  \label{eq:x}
  \alpha(\phi(W(f)))=\pi(W(f)),\ \forall f\in S, \\
  \mbox{(``primitives'')}\ \ \ \ \ \ \ \ \ \ \ \ \ \ \nonumber
\end{eqnarray}
\begin{eqnarray}
\label{eq:y}
\beta(\omega)(\alpha(A))=\omega(A),\ \forall\ \mbox{states}\ 
\omega,\ \forall\ \mbox{observables}\ A. \\ 
\mbox{(``theorems'')}\  \ \ \ \ \ \ \ \ \ \ \ \ \ \ \ \ \ \ \ \ \ \ \ \  \nonumber
\end{eqnarray} 
Of course, the notion of equivalence we obtain depends on how we
construe the phrases ``physical observables of a representation
$\pi$'' and ``physical states of a representation $\pi$''.  According
to a conservative rendering of observables, only the self-adjoint
elements of the Weyl algebra $\pi (\alg{W})$ are genuine physical
observables of the representation $\pi$.  (More generally, an
unbounded operator on $\hil{H}_{\pi}$ is a physical observable only if
all of its bounded functions lie in $\pi (\alg{W})$.)  On the other
hand, a liberal rendering of observables considers all self-adjoint
operators in the weak closure $\pi (\alg{W})^{-}$ of $\pi (\alg{W})$
as genuine physical observables.  (More generally, those unbounded
operators whose bounded functions lie in $\pi (\alg{W})^{-}$ should be
considered genuine physical observables.)  A conservative with respect
to states claims that only those density operator states (i.e., normal
states) of the algebra $\pi (\alg{W})$ are genuine physical states.
On the other hand, a liberal with respect to states claims that all
algebraic states of $\pi (\alg{W})$ should be thought of as genuine
physical states.  We thereby obtain \emph{four distinct} necessary
conditions for physical equivalence, according to whether one is
conservative or liberal about observables, and conservative or liberal
about states.

Arageorgis ([1995], p. 302) and Arageorgis \emph{et al} ([2000], p. 3)
also take the correct notion of physical equivalence in this context
to be intertranslatability.  On the basis of informal discussions
(with rather less supporting argument than one would have liked), they
claim that physical equivalence of representations requires that they
be unitarily equivalent.  (They do not discuss quasi-equivalence.) We
disagree with this conclusion, but there is still substantial overlap
between us.  For instance, with our precise necessary condition for
physical equivalence above, we can now establish the following
elementary result.
 \begin{prop} \label{hi} Under the conservative approach to states, 
   $\phi$ (factorial) and $\pi$ (irreducible) are physically
   equivalent representations of $\alg{W}$ only if they are
   quasi-equivalent.  \end{prop}
 
 \begin{proof} Let $\omega$ be a normal state of $\phi (\alg{W})$.  Then, by 
   hypothesis, $\beta(\omega)$ is a normal state of $\pi (\alg{W})$.
   Define a state $\rho$ on $\alg{W}$ by \begin{equation} \rho
     (A)=\omega (\phi (A)) ,\qquad A\in \alg{W} .\end{equation} Since
   $\omega$ is normal, $\rho \in \mathfrak{F}(\phi )$.  Define a state
   $\rho '$ on $\alg{W}$ by 
\begin{equation}
\rho '(A)=\beta (\omega)(\pi (A)) ,\qquad A\in \alg{W}.\end{equation}
Since $\beta (\omega)$ is normal, $\rho '\in \mathfrak{F}(\pi )$.
Now, conditions (\ref{eq:x}) and
   (\ref{eq:y}) entail that \begin{equation}
\omega (\phi (A))=\beta (\omega)(\alpha (\phi (A)))=\beta
(\omega)(\pi (A)) ,\end{equation}
for any $A=W(f)\in \alg{W}$, and thus $\rho (W(f))=\rho '(W(f))$ for
any $f\in S$.  However, a state of the Weyl algebra is 
uniquely determined (via linearity
   and uniform continuity) by its action on the generators
   $\{W(f):f\in S\}$.  Thus, $\rho =\rho '$ and since $\rho \in 
\mathfrak{F}(\phi)\cap \mathfrak{F}(\pi )$, it follows that $\phi$ and
$\pi$ are quasi-equivalent.
 \end{proof}
 
\noindent With somewhat more work, 
the following result may also be established.\footnote{Our proof in
  the appendix makes rigorous Arageorgis' brief (and insufficient)
  reference to Wigner's symmetry representation theorem in his
  ([1995], p. 302, footnote).}
 \begin{prop} \label{frog} Under the liberal approach to observables, 
 $\phi$ (factorial) and $\pi$ (irreducible) are 
 physically equivalent representations of 
 $\alg{W}$ only if they are 
 quasi-equivalent.  \end{prop}
  
The above results leave only the position of the ``conservative about
observables/liberal about states'' undecided.  However, we claim,
\emph{pace} Arageorgis \emph{et al}, that a proponent of this position
can satisfy conditions (\ref{eq:x}),(\ref{eq:y}) \emph{without}
committing himself to quasi-equivalence of the representations.  Since
he is conservative about observables, Proposition 1 already guarantees
the existence of a bijective mapping $\alpha$ --- in fact, a
*-isomorphism from the whole of $\phi(\alg{W})$ to the whole of
$\pi(\alg{W})$ --- satisfying (\ref{eq:x}).  And if he is liberal
about states, the state mapping $\beta$ need not map any normal state
of $\phi (\alg{W})$ into a normal state of $\pi (\alg{W})$, bypassing
the argument for Proposition \ref{hi}.  Consider, for example, the
liberal who takes \emph{all} algebraic states of $\phi(\alg{W})$ and
$\pi(\alg{W})$ to be physically significant.  Then for any algebraic
state $\omega$ of $\phi(\alg{W})$, the bijective mapping $\beta$ that
sends $\omega$ to the state $\omega\circ\alpha^{-1}$ on $\pi(\alg{W})$
trivially satisfies condition (\ref{eq:y}) even when $\phi$ and $\pi$
are disjoint.
 
Though we have argued that Segal was conservative about observables,
we are not claiming he was a liberal about states.  In fact, Segal
consistently maintained that only the ``regular states'' of the Weyl
algebra have physical relevance ([1961], p. 7; [1967], pp. 120, 132).
A state $\omega$ of $\alg{W}[S,\sigma]$ is called \emph{regular} just
in case the map $f\mapsto\omega(W(f))$ is continuous on all
finite-dimensional subspaces of $S$; or, equivalently, just in case
the GNS representation of $\alg{W}[S,\sigma]$ determined by $\omega$
is regular (Segal [1967], p. 134).  However, note that, unlike
normality of a state, regularity is representation-\emph{independent}.
Taking the set of all regular states of the Weyl algebra to be
physical is therefore still liberal enough to permit satisfaction of
condition (\ref{eq:y}).  For the mapping $\beta$ of the previous
paragraph trivially preserves regularity, insofar as both $\omega$ and
$\omega\circ\alpha^{-1}$ induce the same abstract regular state of
$\alg{W}$.
  
Our verdict, then, is that Segal is not committed to saying only
quasi-equivalent representations can be physically equivalent.  And
this explains why he sees fit to \emph{define} physical equivalence of
representations in such a way that Proposition 1 secures the physical
equivalence of all representations (see Segal [1961], Defn. 1(c)).
(Indeed, Segal regards Proposition 1 as the appropriate generalization
of the Stone-von Neumann uniqueness theorem to infinite-dimensional
$S$.)  One might still ask what the point of passing to a concrete
Hilbert space representation of $\alg{W}$ is if one is going to allow
as physically possible regular states not in the folium of the chosen
representation.  The point, we take it, is that if we are interested
in drawing out the predictions of some particular regular state, such
as the Minkowski vacuum or the Rindler vacuum, then passing to a
particular representation will put at our disposal all the standard
analytical techniques of Hilbert space quantum mechanics to facilitate
calculations in that particular state.\footnote{In support of not
  limiting the physical states of the Weyl algebra to any one
  representation's folium, one can also cite the cases of
  non-unitarily implementable dynamics discussed by Arageorgis
  \emph{et al} ([2000]) in which dynamical evolution occurs between
  regular states that induce disjoint GNS representations.  In such
  cases, it would hardly be coherent to maintain that regular states
  \emph{dynamically accessible to one another} are not physically
  co-possible.}
 
Haag \& Kastler ([1964], p. 852) and Robinson ([1966], p.  488) have
argued that \emph{by itself} the \emph{weak} equivalence of all
representations of the Weyl algebra entails their physical
equivalence.\footnote{Indeed, the term ``physical equivalence'' is
  often used synonomously with weak equivalence; for example, by Emch
  ([1972], p. 108), who, however, issues the warning that ``we should
  be seriously wary of semantic extrapolations'' from this usage.
  Indeed!}  Their argument starts from the fact that, by measuring the
expectations of a finite number of observables $\{A_{i}\}$ in the Weyl
algebra, each to a finite degree of accuracy $\epsilon$, we can only
determine the state of the system to within a weak* neighborhood. But
by Fell's density theorem, states from the folium of \emph{every}
representation lie in this neighborhood.  So for all practical
purposes, we can never determine which representation is the
physically ``correct'' one and they all, in some (as yet,
unarticulated!) sense, carry the same physical content.  And as a
corollary, choosing a representation is simply a matter of convention.
 
Clearly the necessary condition for physical equivalence we have
proposed constitutes a very different notion of equivalence than weak
equivalence, so we are not disposed to agree with this argument.
Evidently it presupposes that only the observables in the Weyl algebra
itself are physically significant, which we have granted \emph{could}
be grounded in operationalism.  However, there is an additional layer
of operationalism that the argument must presuppose: scepticism about
the physical meaning of postulating an \emph{absolutely precise} state
for the system.  If we follow this scepticism to its logical
conclusion, we should instead think of physical states of the Weyl
algebra as represented by weak* neighborhoods of algebraic states.
What it would then mean to falsify a state, so understood, is that
some finite number of expectation values measured to within finite
accuracy are found to be incompatible with all the algebraic states in
some proposed weak* neighborhood.  Unfortunately, no particular
``state'' in this sense can ever be fully empirically adequate, for
any hypothesized state ($=$ weak* neighborhood) will be subject to
constant revision as the accuracy and number of our experiments
increase.  We agree with Summers [1998] that this would do irreparable
damage to the predictive power of the theory --- damage that can only
be avoided by maintaining that there is a correct algebraic state.
 
We do not, however, agree with Summers' presumption (tacitly endorsed
by Arageorgis \emph{et al} [2000]) that we not only need the correct
algebraic state, but ``\ldots the correct state \emph{in the correct
  representation}'' ([2000], p. 13; italics ours).  This added remark
of Summers' is directed against the conventionalist corollary to
Fell's theorem. Yet we see nothing in the point about predictive power
that privileges any particular representation, not even the GNS
representation of the predicted state.  We might well have good reason
to deliberately choose a representation in which the \emph{precise}
algebraic state predicted is not normal. (For example, Kay [1985] does
exactly this, by ``constructing'' the Minkowski vacuum as a thermal
state in the Rindler quantization.)  The role Fell's theorem plays is
then, at best, methodological.  All it guarantees is that when we
calculate with density operators in our chosen represention, we can
always get a reasonably good indication of the predictions of
\emph{whatever} precise algebraic state we have postulated for the
system.
 
So much for the conservative stance on observables.  An interpreter
of quantum field theory is not likely to find it
attractive, if only because none of the observables that have any
chance of underwriting the particle concept lie in the Weyl algebra.
But suppose, as interpreters, we adopt the liberal approach to observables.  Does the
physical inequivalence of disjoint representations entail their
incompatibility, or even incommensurability?  By this, we do not mean
to conjure up Kuhnian thoughts about incommensurable ``paradigms'',
whose proponents share no methods to resolve their disputes.  Rather,
we are pointing to the (more Feyerabendian?)  possibility of an
unanalyzable shift in meaning between disjoint representations as a
consequence the fact that the concepts (observables and/or states) of
one representation are not wholly definable or translatable in terms
of those of the other.
 
One might think of neutralizing this threat by viewing disjoint
representations as sub-theories or models of a more general theory
built upon the Weyl algebra.  Consider the analogy of \emph{two
  different} classical systems, modelled, say, by phase spaces of
different dimension.  Though not physically equivalent, these models
hardly define incommensurable theories insofar as they share the
characteristic kinematical and dynamical features that warrant the
term ``classical''.  Surely the same could be said of disjoint
representations of the Weyl algebra?
 
Alas, there is a crucial disanalogy.  In the case of the Minkowski and
Rindler representations, physicists freely switch between them to
describe the quantum state of the \emph{very same} ``system'' --- in
this case, the quantum field in a fixed region of spacetime (see,
e.g., Unruh and Wald [1984] and Wald [1994], Sec. 5.1). And, as we
shall see later, the weak closures of these representations provide
physically inequivalent descriptions of the particle content in the
region.  So it is tempting to view this switching back and forth
between disjoint representations as conceptually incoherent
(Arageorgis [1995], p. 268), and to see the particle concepts
associated to the representations as not just different, but outright
incommensurable (Arageorgis \emph{et al} [2000]).
  
We shall argue that this view, tempting as it is, goes too far.  For
suppose we \emph{do} take the view that the observables affiliated to
the von Neumann algebras generated by two disjoint representations
$\phi$ and $\pi$ simply represent different physical aspects of the same
physical system.  If we are also liberal about states (not restricting
ourselves to any one representation's folium), then it is natural to
ask what implications a state $\omega$ of our system, that happens to
be in the folium of $\phi$, has for the observables in
$\pi(\alg{W})''$.  In many cases, it is possible to extract a definite
answer.
  
In particular, any abstract state $\omega$ of $\alg{W}$ gives rise to
a state on $\pi(\alg{W})$, which may be extended to a state on the
weak closure $\pi(\alg{W})''$ (KR [1997], Thm.  4.3.13).  The only
catch is that unless $\omega \in \mathfrak{F}(\pi )$, this extension
will not be unique.  For, only normal states of $\pi (\alg{W})$
possess sufficiently nice continuity properties to ensure that their
values on $\pi(\alg{W})$ uniquely fix their values on the weak-closure
$\pi(\alg{W})''$ (see KR [1997], Thm. 7.1.12).  \emph{However}, it may
happen that all extensions of $\omega$ agree on the expectation value
they assign to a \emph{particular observable} affiliated to
$\pi(\alg{W})''$. This is the strategy we shall use to make sense of
assertions such as ``The Minkowski vacuum in a (Rindler) spacetime
wedge is full of Rindler quanta'' (cf., e.g., DeWitt [1979a]).  The
very fact that such assertions can be made sense of \emph{at all}
takes the steam out of claims that disjoint representations are
incommensurable.  Indeed, we shall ultimately argue that this shows
disjoint representations should not be treated as \emph{competing}
``theories'' in the first place.
 
\section{Constructing representations}
We now explain how to construct ``Fock representations'' of the CCRs.
In sections \textbf{3.1} and \textbf{3.2} we
 show how this construction depends on one's choice of
preferred timelike motion in Minkowski spacetime.  In section 
\textbf{3.3}, we show that alternative choices of preferred
timelike motion can result in unitarily inequivalent --- indeed,
disjoint --- representations.

\subsection{First Quantization (``Splitting the Frequencies'')}
The first step in the quantization scheme consists in turning the
classical phase space $(S,\sigma)$ into a quantum-mechanical ``one
particle space'' --- i.e., a Hilbert space.  \emph{The non-uniqueness
of the quantization scheme comes in at this very first step.}  

Depending on our choice of preferred timelike motion, we will have a
one-parameter group $T_{t}$ of linear mappings from $S$ onto $S$
representing the evolution of the classical system in time.  The flow
$t\mapsto T_{t}$ should also preserve the symplectic form.  A
bijective real-linear mapping $T:S\mapsto S$ is called a
\emph{symplectomorphism} just in case $T$ preserves the symplectic
form; i.e., $\sigma (Tf,Tg)=\sigma (f,g)$ for all $f,g\in S$.

We say that $J$ is a \emph{complex structure} for $(S,\sigma )$ just
in case 
\begin{enumerate} 
\item $J$ is a symplectomorphism, 
\item $J^{2}=-I$,
\item $\sigma (f,Jf)> 0 , \qquad 0\not=f \in S$.
\end{enumerate}  
Relative to a complex structure $J$, we may extend the scalar
multiplication on $S$ to complex numbers; viz., take
multiplication by $a+bi$ as given by $a+bi:=af+bJf\in S$.  We may also define an
inner product $(\cdot ,\cdot )_{J}$ on the resulting complex vector
space by setting
\begin{equation} \label{eq:ip}
(f,g)_{J}:= \sigma (f,Jg)+i\sigma (f,g) , \qquad f,g\in S .\end{equation}
We let $\hil{S}_{J}$ denote the Hilbert space that results when we
equip $(S,\sigma)$ with the extended scalar multiplication 
and inner product $(\cdot ,\cdot )_{J}$.    

A symplectomorphism $T$ is (by assumption) a real-linear operator on
$S$.  However, it does not automatically follow that $T$ is a
\emph{complex}-linear operator on $\hil{S}_{J}$, since $T(if)=i(Tf)$
may fail.  If, however, $T$ commutes with $J$, then $T$ will be a
complex-linear operator on $\hil{S}_{J}$, and it is easy to see that
$(Tf,Tg)_{J}=(f,g)_{J}$ for all $f,g\in \hil{S}_{J}$, so $T$ would in fact be 
unitary.  Accordingly, we say that a group $T_{t}$
of symplectomorphisms on $(S,\sigma )$ is \emph{unitarizable} relative
to $J$ just in case $[J,T_{t}]=0$ for all $t\in \mathbb{R}$.  

If
$T_{t}$ is unitarizable and $t\mapsto T_{t}$ is weakly continuous,
so that we have $T_{t}=e^{itH}$ (by Stone's theorem), we say that $T_{t}$ has
\emph{positive energy} just in case $H$ is a positive operator.  In
general, we say that $(\hil{H},U_{t})$ is a \emph{quantum one
  particle system} just in case $\hil{H}$ is a Hilbert space and
$U_{t}$ is a weakly continuous one-parameter unitary group on
$\hil{H}$ with positive energy.  Kay ([1979]) proved:

\begin{prop}  Let $T_{t}$ be a one-parameter group of
  symplectomorphisms of $(S,\sigma )$.  If there is a complex
  structure $J$ on $(S,\sigma )$ such that $(\hil{S}_{J},T_{t})$ is a
  quantum one particle system, then $J$ is unique.  \label{kay} \end{prop}
  
  \noindent Physically, the time translation group 
  $T_{t}$ determines a 
  natural decomposition (or ``splitting'') of the solutions of the 
  relativistic wave equation we are quantizing into those 
  that oscillate with purely positive and with purely negative frequency 
  with respect to the motion.  This has the effect of uniquely fixing a choice of 
  $J$, and the Hilbert space 
  $\hil{S}_{J}$ then provides a representation of the positive frequency solutions 
  alone.\footnote{For more physical details, see Fulling ([1972], Secs. VIII.3,4) and Wald ([1994], 
  pp. 41-2, 63, 111).}

\subsection{Second Quantization (Fock space)}
Once we have a fixed complex structure $J$ on $(S,\sigma )$, the
``second quantization'' procedure yields a unique representation $(\pi
,\hil{H}_{\pi})$ of the Weyl algebra $\alg{W}[S,\sigma ]$.

Let $\hil{H}^{n}$ denote the $n$-fold symmetric tensor product of
$\hil{S}_{J}$ with itself.  That is, using $\hil{S}_{J}^{n}$ to 
denote $\hil{S}_{J}\otimes \cdots
\otimes \hil{S}_{J}$ ($n$ times), 
 $\hil{H}^{n}=P_{+}(\hil{S}_{J}^{n})$ where $P_{+}$ is the projection onto the
symmetric subspace.  Then we define a Hilbert space
\begin{equation} \hil{F}(\hil{S}_{J}):=\mathbb{C}\oplus \hil{H}^{1}\oplus
  \hil{H}^{2}\oplus \hil{H}^{3} \oplus \cdots ,\end{equation} 
called the \emph{bosonic Fock space over} $\hil{S}_{J}$.  Let \begin{equation}
\Omega := 1\oplus 0 \oplus 0 \oplus \cdots ,\end{equation}
denote the privileged ``Fock vacuum'' state in $\hil{F}(\hil{S}_{J})$. 

Now, we define creation and annihilation
operators on $\hil{F}(\hil{S}_{J})$ in the 
usual way.  For any fixed $f\in S$, we first consider the unique bounded linear 
extensions of the mappings 
$a^{*}_{n}(f):\hil{S}_{J}^{n-1}\rightarrow \hil{S}_{J}^{n}$ and 
$a_{n}(f):\hil{S}_{J}^{n}\rightarrow \hil{S}_{J}^{n-1}$ defined by the following 
actions on product vectors 
\begin{equation}
  a^{*}_{n}(f)(f_{1}\otimes\cdots\otimes f_{n-1})=
  f\otimes f_{1}\otimes\cdots\otimes f_{n-1},
  \end{equation}
  \begin{equation}
  a_{n}(f)(f_{1}\otimes\cdots\otimes f_{n})=
  (f,f_{1})_{J}\ f_{2}\otimes\cdots\otimes f_{n}.
  \end{equation}
We then define the \emph{unbounded} creation and annihilation 
operators on $\hil{F}(\hil{S}_{J})$ by
\begin{equation}
  a^{*}(f):=a^{*}_{1}(f)\oplus \sqrt{2} P_{+}a^{*}_{2}(f)\oplus \sqrt{3} 
  P_{+}a^{*}_{3}(f)\oplus\cdots,
  \end{equation}
\begin{equation}
  a(f):=0\oplus a_{1}(f)\oplus \sqrt{2} a_{2}(f)\oplus \sqrt{3} a_{3}(f)\oplus\cdots.
  \end{equation}
  (Note that the mapping $f\mapsto a^{*}(f)$ is linear while $f\mapsto 
  a(f)$ is \emph{anti}-linear.) 
  
  As the definitions and notation suggest, $a^{*}(f)$ and $a(f)$ are
  each other's adjoint, $a^{*}(f)$ is the creation operator for a
  particle with wavefunction $f$, and $a(f)$ the corresponding
  annihilation operator.  The unbounded self-adjoint operator
  $N(f)=a^{*}(f)a(f)$ represents the number of particles in the field
  with wavefunction $f$ (unbounded, because we are describing bosons
  to which no exclusion principle applies).  Summing $N(f)$ over any
  $J$-orthonormal basis of wavefunctions in $\hil{S}_{J}$, we obtain
  the \emph{total} number operator $N$ on $\hil{F}(\hil{S}_{J})$,
  which has the form
\begin{equation} N=0\oplus 1\oplus
  2\oplus 3 \oplus \cdots .\end{equation} 

Next, we define the self-adjoint ``field operators''
 \begin{equation} \label{eq:star} \Phi (f):= 2^{-1/2}(a^{*}(f)+a(f)) , \qquad f\in S.
\end{equation}
(In heuristic 
discussions of free quantum field theory, these are normally encountered as 
``operator-valued solutions'' $\Phi(x)$ to a relativistic field 
equation at some fixed time.  However, if
 we want to associate a properly defined self-adjoint field operator with the 
 spatial point $x$, we must consider a neighborhood of $x$, and an operator 
 of form $\Phi(f)$, where the ``test-function'' $f\in S$ has support in 
 the neighborhood.\footnote{The picture of a quantum field as an 
 operator-valued \emph{field} --- or, as Teller ([1995], Ch. 5) aptly puts 
 it, a field 
 of ``determinables'' --- unfortunately, has no mathematically rigorous 
 foundation.})
Defining the unitary operators
\begin{equation} \label{eq:reg}
\pi(W(tf)):=\exp (it\Phi (f)), \qquad t\in \mathbb{R},\ f\in S,
\end{equation}
it can then be verified (though it is not trivial) that the
$\pi(W(f))$ satisfy the Weyl form of the CCRs. In fact, the mapping
$W(f)\mapsto \pi (W(f))$ gives an irreducible regular representation
$\pi$ of $\alg{W}$ on $\hil{F}(\hil{S}_{J})$.
  
We also have 
\begin{equation} \langle \Omega ,\pi(W(f))\Omega \rangle 
=e^{-(f,f)_{J}/4}, \qquad f\in S.  \label{eq:vacuum}
\end{equation} (We shall always distinguish the inner product of 
$\hil{F}(\hil{S}_{J})$ from that of $\hil{S}_{J}$ by using angle brackets.) 
The vacuum vector $\Omega \in \hil{F}(\hil{S}_{J})$ defines an
abstract regular state $\omega _{J}$ of $\alg{W}$ via $\omega_{J} (A):=
\langle \Omega ,\pi (A)\Omega \rangle$ for all $A\in
  \alg{W}$.  Since the action of $\pi (\alg{W})$ on
$\hil{F}(\hil{S}_{J})$ is irreducible, $\{ \pi (A)\Omega :A\in \alg{W}
\}$ is dense in $\hil{F}(\hil{S}_{J})$ (else its closure would be a
non-trivial subspace invariant under all operators in $\pi
(\alg{W})$).  Thus, the Fock representation of $\alg{W}$ on
$\hil{F}(\hil{S}_{J})$ is unitarily equivalent to the GNS representation of $\alg{W}$
determined by the pure state $\omega _{J}$.  

In sum, a complex structure $J$ on $(S,\sigma )$ gives rise to an
abstract vacuum state $\omega _{J}$ on $\alg{W}[S,\sigma ]$ whose GNS
representation $(\pi _{\omega_{J}},\hil{H}_{\omega_{J}},\Omega
_{\omega_{J}})$ is just the standard Fock vacuum representation
$(\pi,\hil{F}(\hil{S}_{J}),\Omega)$.  Note also that inverting Eqn.
(\ref{eq:star}) yields
\begin{equation} \label{eq:two}
  a^{*}(f)= 2^{-1/2}(\Phi (f)-i\Phi (if)),\ a(f)= 2^{-1/2}(\Phi (f)+i\Phi 
  (if)),\ f\in S.
  \end{equation}
  Thus, we could just as well have arrived at the Fock representation of $\alg{W}$ 
  ``abstractly'' by \emph{starting} with the pure regular state $\omega 
  _{J}$ on $\alg{W}[S,\sigma]$ as our proposed vacuum,  
  exploiting its regularity
  to guarantee the existence of field operators $\{\Phi (f):f\in S\}$ 
  acting on 
  $\hil{H}_{\omega_{J}}$, and then using Eqns. (\ref{eq:two}) 
  to \emph{define} $a^{*}(f)$ and 
  $a(f)$ (and, from thence, the number operators $N(f)$ and $N$).
  
 There is a natural way to construct operators on
$\hil{F}(\hil{S}_{J})$ out of operators on the 
one-particle space $\hil{S}_{J}$, using the \emph{second 
quantization map} $\Gamma$ and its ``derivative'' $d\Gamma$.  Unlike the 
representation map $\pi$, 
the operators on $\hil{F}(\hil{S}_{J})$ in the range of $\Gamma$ and 
$d\Gamma$ do not ``come from'' 
$\alg{W}[S,\sigma]$, but rather $\mathbf{B}(\hil{S}_{J})$.  Since the latter 
depends on 
how $S$ was complexified, we cannot expect second quantized observables 
to be
representation-independent.   

To define $d\Gamma$, first let
$H$ be a self-adjoint (possibly unbounded) operator on $\hil{S}_{J}$.
We define $H_{n}$ on $\hil{H}^{n}$ by setting $H_{0}=0$ and
\begin{equation}
H_{n}(P_{+}(f_{1}\otimes \cdots \otimes f_{n}))=P_{+}\left( \sum
  _{i=1}^{n} f_{1}\otimes f_{2}\otimes \cdots \otimes Hf_{i} \otimes
  \cdots \otimes f_{n}\right) ,\end{equation}
for all $f_{i}$ in the domain of $H$, and then extending by
  continuity.  It then follows that $\oplus _{n\geq 0}H_{n}$ is an
  ``essentially selfadjoint'' operator on $\hil{F}(\hil{S}_{J})$ (see
  BR [1996], p. 8).  We let \begin{equation}
d\Gamma (H):=\overline{ \bigoplus _{n\geq 0}H_{n} } ,\end{equation}
denote the resulting (closed) self-adjoint operator.  The simplest 
example occurs when we take $H=I$, in which case it is easy to see 
that $d\Gamma (H)=N$.  
In stark contrast to this, we have the following.\footnote{Our 
proof in the appendix reconstructs the argument briefly sketched in Segal ([1959], 
p. 12).} 

\begin{prop} When $S$ is infinite-dimensional, $\pi(\alg{W}[S,\sigma])$
contains no non-trivial bounded functions of the total number operator
  on $\hil{F}(\hil{S}_{J})$.   \label{segal} \end{prop}
  
\noindent In particular, $\pi(\alg{W})$ does not contain any of the spectral 
projections of $N$.  Thus, while the \emph{conservative} about
observables is free to refer to the abstract state $\omega_{J}$ of
$\alg{W}$ as a ``vacuum'' state, he cannot use that language to
underwrite the claim that $\omega_{J}$ is a state of ``no particles''!

To define $\Gamma$, let $U$ be a unitary operator on $\hil{S}_{J}$.  Then
$U_{n}=P_{+}(U\otimes \cdots \otimes U)$ is a unitary operator on
$\hil{H}^{n}$.  We define the unitary operator $\Gamma (U)$ on
$\hil{F}(\hil{S}_{J})$ by
\begin{equation} \Gamma (U):=\bigoplus _{n\geq 0}U_{n} .\end{equation}
If $U_{t}=e^{itH}$ is a weakly continuous unitary group on
$\hil{S}_{J}$, then $\Gamma (U_{t})$ is a weakly continuous group on
$\hil{F}(\hil{S}_{J})$, and we have
\begin{equation} \Gamma (U_{t})=e^{itd\Gamma (H)} .\end{equation}
In particular, the one-particle 
evolution $T_{t}=e^{itH}$ that was used to fix $J$
``lifts''  to a field evolution given by $\Gamma (T_{t})$, where 
$d\Gamma (H)$ represents the energy of the field and has the vacuum $\Omega$ as 
a ground state. 

It can be shown that the representation and second quantization maps
interact as follows: 
\begin{equation}
\pi(W(Uf))=\Gamma (U)^{*}\pi(W(f))\Gamma (U) ,\qquad f\in S,\end{equation}
for any unitary operator $U$ on $\hil{S}_{J}$.  Taking the phase 
transformation $U=e^{it}I$, it follows that
\begin{equation} \label{eq:forty1}
\pi(W(e^{it}f))=e^{-itN}\pi(W(f))e^{itN} ,\qquad f\in S,\ t\in \mathbb{R}. \end{equation}
Using Eqn. (\ref{eq:vacuum}), it also follows that  
\begin{equation}
\langle \Gamma (U)\Omega ,\pi(W(f))\Gamma (U)\Omega \rangle 
=\langle \Omega ,\pi(W(Uf))\Omega
\rangle =\langle \Omega ,\pi(W(f))\Omega
\rangle .  \end{equation}
Since the states induced by the vectors $\Omega$ and $\Gamma 
(U)\Omega$ are both normal in $\pi$ and agree on $\pi(\alg{W})$, they 
determine the same state of 
$\pi(\alg{W})''=\mathbf{B}(\hil{F}(\hil{S}_{J}))$.  Thus 
$\Omega$ must be an eigenvector of $\Gamma
(U)$ for any unitary operator $U$ on $\hil{S}_{J}$.  In particular, 
the vacuum is invariant under the group
$\Gamma(T_{t})$, and is therefore time-translation invariant. 

\subsection{Disjointness of the Minkowski and Rindler representations}
We omit the details of the construction of the classical phase
space $(S,\sigma )$, since they are largely irrelevant to our
concerns.  The only information we need is that the space $S$ may be
taken (roughly) to be solutions to some relativistic wave equation,
such as the Klein-Gordon equation.  More particularly, $S$ may be
taken to consist of pairs of smooth, compactly supported functions on
$\mathbb{R}^{3}$: one function specifies the values of the field at
each point in space at some initial time (say $t=0$), and the other
function is the time-derivative of the field (evaluated at $t=0$).  If
we then choose a ``timelike flow'' in Minkowski spacetime, we will get
a corresponding flow in the solution space $S$; and, in particular,
this flow will be given by a one-parameter group $T_{t}$ of
symplectomorphisms on $(S,\sigma )$.

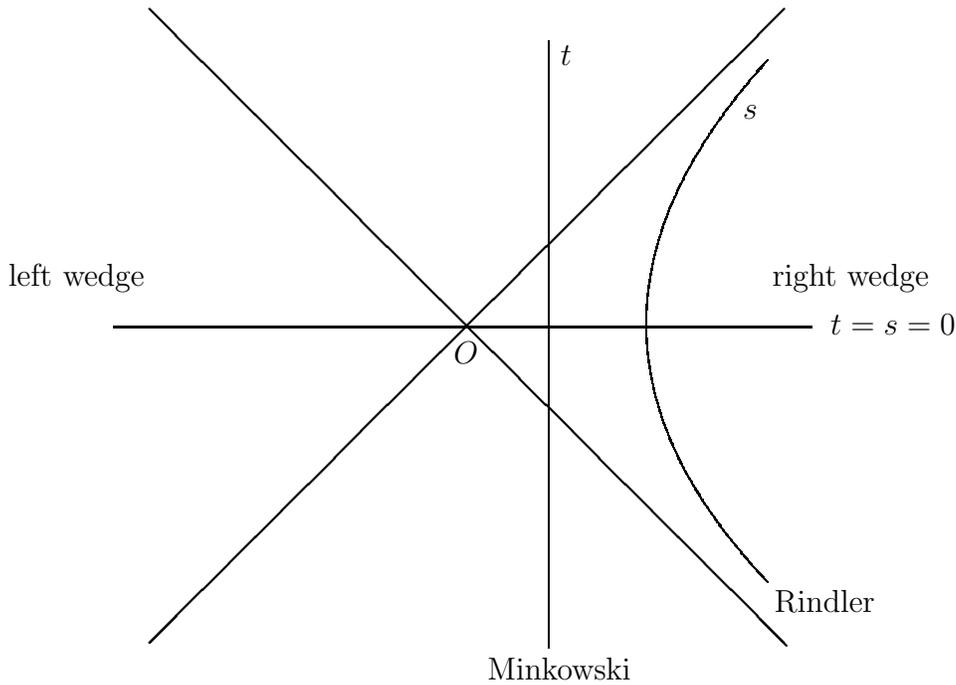
\begin{figure}
\setlength{\unitlength}{0.92pt}
\begin{picture}(320,270)
\thinlines  \qbezier(322,52)(222,157)(322,267)
\put(325,40){Rindler}
\thicklines   \put(68,288){\line(1,-1){262}}
              \put(329,288){\line(-1,-1){261}}
\thinlines    \put(232,275){\line(0,-1){250}}
\put(207,12){Minkowski}
              \put(53,157){\line(1,0){287}}
              \put(10,174){left wedge}
              \put(324,174){right wedge}
              \put(348,154){$t=s=0$}
              \put(312,243){$s$}
              \put(193,142){$O$}
              \put(237,265){$t$}
\end{picture} \caption{Minkowski and Rindler Motions.} 
\label{figure!} \end{figure}

First, consider the group $T_{t}$ of symplectomorphisms of
$(S,\sigma)$ induced by the standard inertial timelike flow. (See
Figure~\ref{figure!}, which suppresses two spatial dimensions.  Note
that it is irrelevant which inertial frame's flow we pick, since they
all determine the same representation of $\alg{W}[S,\sigma ]$ up to
unitary equivalence; see Wald [1994], p. 106.)  It is well-known that
there is a complex structure $M$ on $(S,\sigma )$ such that
$(\hil{S}_{M},T_{t})$ is a quantum one-particle system (see Kay
[1985]; Horuzhy [1988], Ch. 4).  We call the associated pure regular
state $\omega _{M}$ of $\alg{W}[S,\sigma ]$ the \emph{Minkowski vacuum
  state}.  As we have seen, it gives rise via the GNS construction to
a unique Fock vacuum representation $\pi_{\omega_{M}}$ on the Hilbert
space $\hil{H}_{\omega_{M}}=\hil{F}(\hil{S}_{M})$.

Next, consider the group of Lorentz boosts about a given centre point
$O$ in spacetime.  This also gives rise to a one-parameter group
$T_{s}$ of symplectomorphisms of $(S,\sigma )$ (cf.
Figure~\ref{figure!}).  Let $S(\triangleleft)$ be the subspace of $S$
consisting of Cauchy data with support in the right Rindler wedge
($x_{1}>0$); i.e., at $s=0$, both the field and its first derivative
vanish when $x_{1}\leq 0$.  Let
$\alg{W}_{\triangleleft}:=\alg{W}[S(\triangleleft),\sigma ]$ be the
Weyl algebra over the symplectic space $(S(\triangleleft),\sigma )$.
Then, $T_{s}$ leaves $S(\triangleleft)$ invariant, and hence gives
rise to a one-parameter group of symplectomorphisms of
$(S(\triangleleft),\sigma )$.  Kay ([1985]) has shown rigorously that
there is indeed a complex structure $R$ on $(S(\triangleleft),\sigma
)$ such that $(\hil{S}(\triangleleft)_{R},T_{s})$ is a quantum one
particle system.  We call the resulting state $\omega
_{R}^{\triangleleft}$ of $\alg{W}_{\triangleleft}$ the \emph{(right)
  Rindler vacuum state}.  It gives rise to a unique GNS-Fock
representation $\pi_{\omega _{R}^{\triangleleft}}$ of
$\alg{W}_{\triangleleft}$ on $\hil{H}_{\omega
  _{R}^{\triangleleft}}=\hil{F}(\hil{S}(\triangleleft)_{R})$ and,
hence, a quantum field theory for the spacetime consisting of the
right wedge \emph{alone}.  

The Minkowski vacuum state $\omega _{M}$ of $\alg{W}$ also determines a 
state $\omega _{M}^{\triangleleft}$ of 
$\alg{W}_{\triangleleft}$, by restriction (i.e., 
$\omega _{M}^{\triangleleft}:=\omega _{M}|_{\alg{W}_{\triangleleft}}$). 
 Thus, we may apply the GNS
construction to obtain the Minkowski representation $(\pi
_{\omega _{M}^{\triangleleft}},\hil{H}_{\omega _{M}^{\triangleleft}})$ 
of $\alg{W}_{\triangleleft}$.  It can be shown (using the 
``Reeh-Schlieder theorem'' --- see Clifton and Halvorson [2000]) 
that $\omega _{M}^{\triangleleft}$ is a highly mixed state 
(unlike $\omega _{R}^{\triangleleft}$).  Therefore, $\pi
_{\omega _{M}^{\triangleleft}}$ is reducible.  

To obtain a concrete picture of this representation, note that (again,
as a consequence of the ``Reeh-Schlieder theorem'') $\Omega _{\omega
  _{M}}$ is a cyclic vector for the subalgebra $\pi _{\omega
  _{M}}(\alg{W}_{\triangleleft})$ acting on the ``global'' Fock space
$\hil{F}(\hil{S}_{M})$.  Thus, by the uniqueness of the GNS
representation $(\pi_{\omega _{M}^{\triangleleft}},\hil{H}_{\omega
  _{M}^{\triangleleft}})$, it is unitarily equivalent to the
representation $(\pi _{\omega
  _{M}}|_{\alg{W}_{\triangleleft}},\hil{F}(\hil{S}_{M}))$.  It can be
shown that $\pi _{\omega _{M}}(\alg{W}_{\triangleleft})''$ is a factor
(Horuzhy ([1988], Thm. 3.3.4).  Thus, while reducible, $\pi_{\omega
  _{M}^{\triangleleft}}$ is still factorial.
   
Under the liberal approach to observables, the representations 
$\pi_{\omega _{M}^{\triangleleft}}$ (factorial) and 
$\pi_{\omega _{R}^{\triangleleft}}$ (irreducible) provide physically 
inequivalent descriptions of the physics in the right wedge.  

\begin{prop} \label{disjoint1} The Minkowski and Rindler representations of
  $\alg{W}_{\triangleleft}$ are disjoint.  \end{prop}
  
Now let $\triangleright$ denote the left Rindler wedge, and define the
subspace $S(\triangleright)$ of $S$ as $S(\triangleleft)$ was defined
above.  (Of course, by symmetry, Proposition \ref{disjoint1} holds for
$\alg{W}_{\triangleright}$ as well.)  Let
$\alg{W}_{\bowtie}:=\alg{W}[S(\triangleright)\oplus
S(\triangleleft),\sigma ]$ denote the Weyl algebra over the symplectic
space $(S(\triangleright)\oplus S(\triangleleft),\sigma )$.  Then
$\alg{W}_{\bowtie}=\alg{W}_{\triangleright}\otimes
\alg{W}_{\triangleleft}$, and $\omega _{M}^{\bowtie} :=\omega
_{M}|_{\alg{W}_{\bowtie}}$ is \emph{pure} (Kay [1985], Defn., Thm.
1.3(iii)).\footnote{The restriction of $\omega _{M}$ to
  $\alg{W}_{\bowtie}$ is a pure ``quasifree'' state.  Thus, there is a
  complex structure $M'$ on $S(\triangleright)\oplus S(\triangleleft)$
  such that \begin{equation} \omega _{M}(W(f))=\exp (-\sigma
    (f,M'f)/4) =\exp (-\sigma (f,Mf)/4 ) ,\end{equation} for all $f\in
  S(\triangleright)\oplus S(\triangleleft)$ (Petz [1990], Prop.~3.9).
  It is not difficult to see then that $M|_{S(\triangleright)\oplus
    S(\triangleleft)}=M'$ and therefore that $M$ leaves
  $S(\triangleright)\oplus S(\triangleleft)$ invariant.  Hereafter, we
  will use $M$ to denote the complex structure on $S$ as well as its
  restriction to $S(\triangleright)\oplus S(\triangleleft)$.}  The GNS
representation $\omega _{M}^{\bowtie}$ induces is therefore
irreducible, and (again invoking the uniqueness of the GNS
representation) it is equivalent to
$(\pi_{\omega_{M}}|_{\alg{W}_{\bowtie}},\hil{F}(\hil{S}_{M}))$ (since
$\Omega _{\omega _{M}}\in\hil{F}(\hil{S}_{M})$ is a cyclic vector for
the subalgebra $\pi _{\omega _{M}}(\alg{W}_{\bowtie})$ as well).

The tensor product of the pure left and right Rindler vacua
$\omega_{R}^{\bowtie}:=\omega ^{\triangleright}_{R} \otimes \omega
^{\triangleleft}_{R}$ is of course also a pure state of
$\alg{W}_{\bowtie}$.\footnote{More precisely, $\omega
  _{R}^{\triangleleft}$ arises from a complex structure
  $R_{\triangleleft}$ on $S(\triangleleft )$, $\omega
  _{R}^{\triangleright}$ arises from a complex structure
  $R_{\triangleright}$ on $S(\triangleright )$, and $\omega
  _{R}^{\bowtie}$ arises from the complex structure $R_{\triangleright
    }\oplus R_{\triangleleft}$ of $S(\triangleright)\oplus
  S(\triangleleft)$.  When no confusion can result, we will use $R$ to
  denote the complex structure on $S(\triangleright)\oplus
  S(\triangleleft)$ and its restriction to $S(\triangleleft )$.}  It
will induce a GNS representation of the latter on the Hilbert space
$\hil{H}_{\omega _{R}^{\bowtie}}$ given by $\hil{F}(\hil{S}_{R})\equiv
\hil{F}(\hil{S}(\triangleright)_{R})\otimes
\hil{F}(\hil{S}(\triangleleft)_{R})$. It is not difficult to show that
$\omega_{R}^{\bowtie}$ and $\omega _{M}^{\bowtie}$, both now
irreducible, are also disjoint.

\begin{prop} \label{disjoint2} The Minkowski and Rindler representations of 
$\alg{W}_{\bowtie}$ are disjoint.  \end{prop}

In our final main section we shall discuss the conceptually
problematic implications that the $M$-\emph{vacuum} states
$\omega_{M}^{\bowtie}$ and $\omega_{M}^{\triangleleft}$ have for the
\emph{presence} of $R$-quanta in the double and right wedge spacetime
regions.  However, we note here an important difference between
Rindler and Minkowski observers.

The total number of $R$-quanta, according to a Rindler observer
confined to the left (resp., right) wedge, is represented by the
number operator $N_{\triangleright}$ (resp., $N_{\triangleleft}$) on
$\hil{F}(\hil{S}(\triangleright)_{R})$ (resp.,
$\hil{F}(\hil{S}(\triangleleft)_{R}$)).  However, because of the
spacelike separation of the wedges, no single Rindler observer has
access, even in principle, to the expectation value of the ``overall''
total Rindler number operator $N_{R}=N_{\triangleright}\otimes
I+I\otimes N_{\triangleleft}$ acting on
$\hil{F}(\hil{S}(\triangleright)_{R})\otimes
\hil{F}(\hil{S}(\triangleleft)_{R})$.

The reverse is true for a Minkowski observer.  While she has access,
at least in principle, to the total number of $M$-quanta operator
$N_{M}$ acting on $\hil{F}(\hil{S}_{M})$, $N_{M}$ is a purely global
observable that does not split into the sum of two separate number
operators associated with the left and right wedges (as a general
consequence of the ``Reeh-Schlieder theorem'' --- see Redhead [1995]).
In fact, since the Minkowski complex structure $M$ is an
``anti-local'' operator (Segal and Goodman [1965]), it fails to leave
either of the subspaces $S(\triangleright)$ or $S(\triangleleft)$
invariant, and it follows that no $M$-quanta number operator is
affiliated with $\pi_{\omega
  _{M}^{\triangleleft}}(\alg{W}_{\triangleleft})''$.\footnote{See
  Halvorson [2000b] for further details and a critical analysis of
  different approaches to the problem of particle localization in
  quantum field theory.}  Thus, even a liberal about observables must
say that a Minkowski observer with access only to the right wedge does
not have the capability of counting $M$-quanta.   

So, while it might be sensible to ask for the probability in state 
$\omega_{M}^{\triangleleft}$ that a 
Rindler observer detects particles in the right wedge, 
it is \emph{not} sensible to ask, conversely, for the probability in state 
$\omega_{R}^{\triangleleft}$ that a Minkowski observer will detect 
particles in the right wedge.  Note also that since $N_{M}$ is a purely 
global observable (i.e., there is no sense to be made of ``the number of 
Minkowski quanta in a bounded spatial or spacetime region''),
what a Minkowski observer 
might \emph{locally} detect with a ``particle detector'' (over an 
extended, but finite, interval of time) can at best 
give an approximate indication of the global Minkowski particle content of 
the field.  

\section{Minkowski versus Rindler Quanta}

We have seen that a Rindler observer will construct ``his quantum
field theory'' of the right wedge spacetime region differently from a
Minkowski observer.  He will use the complex structure $R$ picked out
uniquely by the boost group about $O$, and build up a representation
of $\alg{W}_{\triangleleft}$ on the Fock space
$\hil{F}(\hil{S}(\triangleleft)_{R})$.  However, suppose that the
state of $\alg{W}_{\triangleleft}$ is the state
$\omega_{M}^{\triangleleft}$ of \emph{no} particles (globally!)
according to a Minkowski observer.  What, if anything, will our
Rindler observer say about the particle content in the right wedge?
And does \emph{this} question even make sense?
  
We shall argue that it does, notwithstanding the disjointness of the 
Minkowski and Rindler representations.  And the answer is surprising.  Not only does a Rindler 
observer have a nonzero chance of detecting the presence of 
$R$-quanta.  In section \textbf{4.2} we shall show that if our 
Rindler observer were able to build a detector sensitive to the 
\emph{total} number of $R$-quanta in the right wedge, he would always 
find that the probability of an \emph{infinite} total number is 
\emph{one}!  

We begin in section \textbf{4.1} by discussing the paradox of 
observer-dependence of particles to which such results  
lead.  In particular, we shall criticize Teller's ([1995,1996]) 
resolution of this paradox.  Later, in section \textbf{4.3}, we shall 
also criticize the arguments of 
Arageorgis [1995] and Arageorgis et al [2000] for the 
incommensurability of inequivalent particle concepts, and argue, 
instead, for their complementarity (in \emph{support} of Teller).   

\subsection{The Paradox of the Observer-Dependence of Particles}

Not surprisingly, physicists initially found a Rindler observer's
ability to detect particles in the Minkowski vacuum paradoxical (see
R\"{u}ger [1989], p. 571; Teller [1995], p. 110).  After all,
particles are the sorts of things that are either there or not there,
so how could their presence depend on an observer's state of motion?

One way to resist this paradox is to reject from the outset the
physicality of the Rindler representation, thereby withholding bona
fide particle status from Rindler quanta.  For instance, one could be
bothered by the fact the Rindler representation cannot be globally
defined over the whole of Minkowski spacetime, or that the
one-particle Rindler Hamiltonian lacks a mass gap, allowing an
arbitrarily large number of $R$-quanta to have a fixed finite amount
of energy (``infrared divergence'').  Arageorgis ([1995], Ch. 6) gives
a thorough discussion of these and other ``pathologies'' of the
Rindler representation.\footnote{See also, more recently,
  Belinski\u{i} [1997], Fedotov \emph{et al}. [1999], and Nikoli\'{c}
  [2000].}  In consequence, he argues that the phenomenology
associated with a Rindler observer's ``particle detections'' in the
Minkowski vacuum ought to be explained entirely in terms of
observables affiliated to the Minkowski representation (such as
garden-variety Minkowski vacuum fluctuations of the local field
observables).
 
This is not the usual response to the paradox of observer-dependence.
R\"{u}ger [1989] has characterized the majority of physicists'
responses in terms of the \emph{field approach} and the \emph{detector
  approach}.  Proponents of the field approach emphasize the need to
forfeit particle talk at the fundamental level, and to focus the
discussion on measurement of local field quantities.  Those of the
detector approach emphasize the need to relativize particle talk to
the behaviour of concrete detectors following specified world-lines.
Despite their differing emphases, and the technical difficulties in
unifying these programs (well-documented by Arageorgis [1995]),
neither eschews the Rindler representation as unphysical, presumably
because of its deep connections with quantum statistical mechanics and
blackhole thermodynamics (Sciama \emph{et al} [1981]).  Moreover,
pathological or not, it remains of philosophical interest to examine
the consequences of taking the Rindler representation seriously ---
just as the possibility of time travel in general relativity admitted
by certain ``pathological'' solutions to Einstein's field equations is
of interest.  And it is remarkable that there should be \emph{any}
region of Minkowski spacetime that admits two physically inequivalent
quantum field descriptions.
 
 Teller ([1995,1996]) has recently offered his own resolution of 
 the paradox.  We reproduce below the relevant portions of his discussion in Teller ([1995], 
 p. 111).  
 However, note that he does not distinguish between left 
 and right Rindler observers, $|0;M\rangle$ refers, 
in our notation, to the Minkowski vacuum vector 
$\Omega_{\omega_{M}}\in\hil{F}(\hil{S}_{M})$, and 
$|1,0,0,\ldots\rangle_{M}$ (resp., $|1,0,0,\ldots\rangle_{R}$) 
is a one-particle state 
$0\oplus f\oplus 0\oplus 0\oplus\cdots\in\hil{F}(\hil{S}_{M})$  
(resp., $\in\hil{F}(\hil{S}_{R})$).     
\begin{quote}
  \ldots Rindler raising and lowering operators are expressible as
  superpositions of the Minkowski raising and lowering operators, and
  states with a definite number of Minkowski quanta are superpositions
  of states with different numbers of Rindler quanta. In particular,
  $|0;M\rangle$ is a superposition of Rindler quanta states, including
  states for arbitrarily large numbers of Rindler quanta. In other
  words, $|0;M\rangle$ has an exact value of zero for the Minkowski
  number operator, and is simultaneously highly indefinite for the
  Rindler number operator.\smallskip

...In $|0;M\rangle$ there is \textit{no} definite number of Rindler
quanta. There is only a propensity for detection of one or another number of
Rindler quanta by an accelerating detector. A state in which a quantity has
no exact value is one in which no values for that quantity are definitely,
and so actually, exemplified. Thus in $|0;M\rangle$ no Rindler
quanta actually occur, so the status of $|0;M\rangle$ as a state
completely devoid of quanta is not impugned.\smallskip

To be sure, this interpretive state of affairs is surprising.  
To spell it out one step further, in $|1,0,0,\ldots\rangle_{M}$ there is 
one actual Minkowski quantum, no actual Rindler quanta, and all sorts 
of propensities for manifestation of Rindler quanta, among other 
things.   In $|1,0,0,\ldots\rangle_{R}$ the same comment applies with 
the role of Minkowski and Rindler reversed.  It turns out
that there are various kinds of quanta, and a state in which one kind of
quanta actually occurs is a state in which there are only propensities for
complementary kinds of quanta. Surprising, but perfectly consistent and
coherent. 
\end{quote}

\noindent Teller's point is that $R$-quanta only exist (so to speak)
potentially in the $M$-vacuum, not actually.  Thus it is 
still an invariant observer-independent fact that there are no \emph{actual quanta} in 
the field, and the paradox evaporates.  Similarly for Minkowski 
states of one or more particles as seen by Rindler observers.  There 
is the same definite number of \emph{actual quanta} for all observers.  
Thus, since actual particles are the ``real stuff'', the real stuff 
\emph{is} invariant!  

Notice, however, that there is something 
self-defeating in Teller's final concession, urged by advocates of the 
field and detector 
approaches,
that different kinds of quanta need to be 
distinguished.  For if we do draw the distinction sharply, it is no 
longer clear why even the actual presence of $R$-quanta in the $M$-vacuum 
should bother us.  
Teller seems to want to have it both ways: while there are different 
kinds of quanta, there is still only one kind of \emph{actual} quanta, and it 
better be invariant.   

Does this invariance really hold?  In one sense, Yes.  Disjointness does not 
prevent us from building Rindler 
creation and annihilation operators on the Minkowski representation 
space $\hil{F}(\hil{S}_{M})$.  We simply need to define Rindler 
analogues, $a_{R}^{*}(f)$  and $a_{R}(f)$, of the Minkowski creation 
and annihilation operators 
via Eqns. (\ref{eq:two}) with $\Phi(Rf)$ in place of $\Phi(if)$ 
($=\Phi(Mf)$) (noting that $f\mapsto a_{R}(f)$ will now be anti-linear 
with respect to the \emph{Rindler} conjugation $R$).   It 
is then easy to see, using (\ref{eq:star}), that
\begin{equation} \label{eq:above}   
a_{R}(f)=2^{-1}[a_{M}^{*}((I+MR)f)+a_{M}((I-MR)f)].
\end{equation}
This linear combination would be trivial if $R=\pm M$.  However, 
we know 
$R\not=M$, and $R=-M$ is ruled out because it is inconsistent with 
both complex structures being positive definite.  
Consequently, $\Omega_{\omega_{M}^{\bowtie}}$ must be a 
nontrivial superposition of 
eigenstates of 
the Rindler number operator $N_{R}(f):=a_{R}^{*}(f)a_{R}(f)$; for an 
easy calculation, using (\ref{eq:above}), reveals that  
\begin{equation} \label{eq:invariant?}
N_{R}(f)\Omega_{\omega_{M}^{\bowtie}}=
2^{-2}[\Omega_{\omega_{M}^{\bowtie}}
+a_{M}^{*}((I-MR)f)a_{M}^{*}((I+MR)f)\Omega_{\omega_{M}^{\bowtie}}],
\end{equation}
which (the presence of the nonzero second term guarantees) is not
 a simply a multiple of $\Omega_{\omega_{M}^{\bowtie}}$.  Thus, 
Teller  would be correct to conclude that the Minkowski vacuum implies dispersion 
in the number operator $N_{R}(f)$.  And the same conclusion would follow if, 
instead, we considered the 
Minkowski creation and annihilation operators as acting on the Rindler 
representation space $\hil{F}(\hil{S}_{R})$.  Since only finitely 
many degrees of freedom are involved, this is guaranteed by the 
Stone-von Neumann theorem.

However, therein lies the rub.  $N_{R}(f)$ merely 
represents the number of $R$-quanta with a specified wavefunction $f$.  What about 
the \emph{total} number of $R$-quanta in the $M$-vacuum (which 
involves \emph{all} degrees of freedom)?  If Teller 
cannot assure us that this too has dispersion, his case for the 
invariance of ``actual quanta'' is left in tatters.  
In his 
discussion, Teller fails to distinguish $N_{R}(f)$ from the total 
number operator $N_{R}$, but the distinction is crucial.  
It is a well-known consequence of the disjointness of $\pi_{\omega_{R}^{\bowtie}}$ and 
$\pi_{\omega_{M}^{\bowtie}}$ that neither representation's total number operator 
is definable on the Hilbert space of the other (BR [1996], Thm. 5.2.14).  Therefore, it is 
literally \emph{nonsense} to speak of $\Omega_{\omega_{M}^{\bowtie}}$ as a 
superposition of eigenstates of $N_{R}$!\footnote{In their review of Teller's [1995] book, 
Huggett and Weingard [1996] question whether Teller's ``quanta interpretation''
of quantum field theory can be implemented in the context of inequivalent 
representations.  However, when they discuss 
Teller's
resolution of the observer-dependence paradox, in terms of mere 
\emph{propensities to display} $R$-quanta in the $M$-vacuum, they write ``This seems all well and 
good'' ([1996], p. 309)!  Their only criticism is 
the obvious one: legitimizing such propensity talk ultimately 
requires a solution to the measurement problem.  Teller's response to their 
review is equally unsatisfactory.  Though he pays 
lip-service to the possibility of inequivalent representations 
([1998], pp. 156-7), he fails to notice 
how inequivalence undercuts his discussion of the paradox.}  
If $x_{n},x_{m}\in \hil{F}(\hil{S}_{R})$ are
eigenstates of $N_{R}$ with eigenvalues $n,m$
respectively, then $x_{n}+x_{m}$ again lies in
$\hil{F}(\hil{S}_{R})$, 
and so is ``orthogonal'' to all eigenstates of the Minkowski number 
operator $N_{M}$ acting on $\hil{F}(\hil{S}_{M})$.  And, indeed, taking infinite sums of Rindler
number eigenstates will again leave us in the folium of the Rindler
representation.  As Arageorgis ([1995], p. 303) has also noted: ``The
Minkowski vacuum state is not a superposition of Rindler quanta
states, despite `appearances'~''.\footnote{Arageorgis presumes 
Teller's discussion is based upon the appearance of the following purely formal 
(i.e., non-normalizable) expression 
for $\Omega_{\omega_{M}^{\bowtie}}$ as a superposition in 
$\hil{F}(\hil{S}_{R})\equiv\hil{F}(\hil{S}(\triangleright)_{R})\otimes
\hil{F}(\hil{S}(\triangleleft)_{R})$ over left (``I'') and right (``II'') Rindler 
modes (Wald [1994], Eqn. (5.1.27)): 
\begin{equation} \nonumber
\prod_{i}\left\{ \sum_{n=0}^{\infty }\exp (-n\pi \omega _{i}/a)\left|
n_{iI}\right\rangle \otimes \left| n_{iII}\right\rangle \right\}\ .
\end{equation} However, it bears mentioning that, as this expression 
suggests: (a) the restriction of 
$\omega_{M}^{\bowtie}$ 
to either $\alg{W}_{\triangleright}$ or $\alg{W}_{\triangleleft}$ is 
indeed mixed; (b)  $\omega_{M}^{\bowtie}$ can be shown rigorously to be an entangled 
state of $\alg{W}_{\triangleright}\otimes\alg{W}_{\triangleleft}$ 
(Clifton and Halvorson [2000]); and (c) the thermal properties of the 
``reduced density matrix'' for either wedge obtained from this formal 
expression can be derived rigorously (Kay [1985]).   In addition, see Propositions 
\ref{teller} and \ref{fulling} below!}   

Yet this point, by itself, does not tell us that Teller's 
discussion cannot be salvaged.  Recall that 
a state $\rho$ is \emph{dispersion-free} on a (bounded)
observable $X$ just in case $\rho (X^{2})=\rho (X)^{2}$.  Suppose, now,
that $Y$ is a possibly \emph{un}bounded observable that is definable in some
representation $\pi$ of $\alg{W}$.  We can then rightly 
say that an algebraic state $\rho$ of $\alg{W}$ \emph{predicts dispersion in} $Y$ just
in case, for \emph{every} extension $\hat{\rho}$ of $\rho$ to
$\pi(\alg{W})''$, $\hat{\rho}$ is \emph{not}
dispersion-free on all bounded functions of $Y$. We then have the 
following result.

\begin{prop}  \label{teller} If $J_{1},J_{2}$ are distinct complex structures on 
$(S,\sigma )$, then $\omega _{J_{1}}$ (resp., $\omega _{J_{2}}$) predicts 
dispersion in $N_{J_{2}}$ (resp., $N_{J_{1}}$). \end{prop}

 \noindent As a consequence, the Minkowski vacuum $\omega_{M}^{\bowtie}$ 
indeed predicts dispersion in the Rindler total number operator $N_{R}$ (and 
in both $N_{\triangleright}\otimes I$ and $I\otimes N_{\triangleleft}$, 
invoking the symmetry between the wedges).
 
 Teller also writes of the Minkowski vacuum as being a superposition 
 of eigenstates of the Rindler number operator with 
 \emph{arbitrary large} eigenvalues.  Eschewing the language of 
 superposition, the idea that there is no finite number of $R$-quanta 
 to which the $M$-vacuum assigns probability one can also be 
 rendered sensible.  The relevant result was first proved by Fulling 
 ([1972], Appendix F; [1989], p. 145): 
 \begin{fulling} Two Fock vacuum representations
  $(\pi, \hil{F}(\hil{H}),\Omega)$ and $(\pi', 
  \hil{F}(\hil{H}'),\Omega')$ of $\alg{W}$ are unitarily equivalent 
  if and only if $\langle \Omega, 
  N'\Omega\rangle<\infty$ (or, equivalently, $\langle \Omega', 
  N\Omega'\rangle<\infty$).  \end{fulling}
  \noindent As stated, this ``theorem'' also fails to make 
  sense, because it is only in the case where the representations are 
  \emph{already} equivalent that  
   the primed total number operator is definable
  on the unprimed representation space and an expression like ``$\langle \Omega, 
  N'\Omega\rangle$'' is well-defined.  (We say more about why this is 
  so in the next section.) However, there \emph{is} a way to understand the 
  expression ``$\langle \Omega,N'\Omega\rangle<\infty$'' (resp., ``$\langle \Omega, 
  N'\Omega\rangle=\infty$'') in a rigorous, non-question-begging way.  
  We can take it to be 
  the claim that all extensions $\hat{\rho}$ of the 
  abstract unprimed vacuum state of $\alg{W}$
  to $\mathbf{B}(\hil{F}(\hil{H}'))$ assign (resp., do \emph{not} assign) $N'$ a finite
  value; i.e., for any such extension, $\sum_{n'=1}^{\infty}\hat{\rho}(P_{n'})n'$ converges 
  (resp., does not converge), 
  where $\{P_{n'}\}$ are the spectral projections of $N'$.  With this 
  understanding, the
  following rigorization of Fulling's ``theorem'' can then be proved.  
  
 \begin{prop}  \label{fulling} A pair of Fock representations
$\pi_{\omega_{J_{1}}}$, $\pi_{\omega_{J_{2}}}$ are unitarily 
equivalent if and only if
$\omega_{J_{1}}$ assigns $N_{J_{2}}$ a finite value (equivalently, 
$\omega_{J_{2}}$ assigns $N_{J_{1}}$ a finite value). 
\end{prop}
 
 \noindent It follows that $\omega_{M}^{\bowtie}$ cannot assign 
 probability one to any finite number of $R$-quanta (and vice-versa, with 
 $R\leftrightarrow M$).
    
 Unfortunately, neither Proposition \ref{teller} or \ref{fulling} is sufficient to rescue 
 Teller's ``actual quanta'' invariance argument, for these propositions give 
 no further information about the shape of the probability distribution 
 that $\omega_{M}^{\bowtie}$ prescribes for $N_{R}$'s eigenvalues.  
 In particular, both propositions 
 are compatible with there being a probability of \emph{one} that 
 \emph{at least} $n>0$ 
 $R$-quanta obtain in the $M$-vacuum, for any $n\in\mathbb{N}$.  If that 
 were the case, Teller 
 would then be forced to withdraw and concede that at least \emph{some}, and 
 perhaps many, Rindler quanta \emph{actually} occur in a state with no 
 actual Minkowski quanta.  In the next section, we shall show that 
 this ---
 Teller's worst nightmare --- is in fact the case.  

\subsection{Minkowski Probabilities for Rindler Number Operators}

We now defend the claim that a Rindler observer will say that there
are actually \emph{infinitely many} quanta while the field is in the
Minkowski vacuum state (or, indeed, in any other state of the
Minkowski folium).\footnote{In fact, this was first proved, in effect,
  by Chaiken [1967].  However his lengthy analysis focussed on
  comparing Fock with non-Fock (so-called ``strange'') representations
  of the Weyl algebra, and the implications of his result for disjoint
  Fock representations based on inequivalent one-particle structures
  seem not to have been carried down into the textbook tradition of
  the subject.  (The closest result we have found is BR ([1996], Thm.
  5.2.14) which we are able to employ as a lemma to recover Chaiken's
  result for disjoint Fock representations --- see the appendix.)}
This result applies more generally to any pair of disjoint regular
representations, at least one of which is the GNS representation of an
abstract Fock vacuum state. We shall specialize back down to the
Minkowski/Rindler case later on.
  
Let $\rho$ be a regular state of $\alg{W}$ inducing the GNS 
representation $(\pi
_{\rho},\hil{H}_{\rho})$, and let $\omega_{J}$ be the abstract vacuum 
state determined by a complex structure $J$ on $(S,\sigma)$.  The 
case we are interested in is, of course, when $\pi
_{\rho}$, $\pi_{\omega_{J}}$ are disjoint.  We first want to show how to define representation-independent 
probabilities in the state $\rho$ for any $J$-quanta number operator 
that ``counts'' the number of quanta with wavefunctions in a fixed 
\emph{finite}-dimensional subspace 
$F\subseteq\hil{S}_{J}$.  (Parts of our exposition 
below follow BR ([1996], pp. 26-30), which may be consulted for further 
details.) 

We know 
that, for any $f\in S$, there exists a
self-adjoint operator $\Phi_{\rho}(f)$ on
$\hil{H}_{\rho}$ such that
\begin{equation} \pi _{\rho} (W(tf))=\exp \left( it\Phi_{\rho}(f)
  \right) , \qquad t\in \mathbb{R}. \end{equation}  
We can also define unbounded annihilation and creation operators on 
$\hil{H}_{\rho}$ for $J$-quanta by
\begin{equation} a_{\rho}(f) := 2^{-1/2}(\Phi_{\rho}(f)+i\Phi
  _{\rho}(Jf)) ,\ 
a^{*}_{\rho}(f):=2^{-1/2}(\Phi_{\rho}(f)-i\Phi_{\rho}(Jf)) .
\end{equation}
Earlier, we denoted these operators by $a_{J}(f)$ and $a^{*}_{J}(f)$.  However, we now 
want to emphasize the representation space upon which they act; and only the single complex structure $J$ 
shall concern us in our general discussion, so there is no possibility of confusion with others. 

Next, define a ``quadratic form'' $n_{\rho}(F):\hil{H}_{\rho}\mapsto\mathbb{R}^{+}$. 
 The domain of $n_{\rho}(F)$
is \begin{equation} D(n_{\rho}(F)):=\bigcap _{f\in F} D(a_{\rho}(f))
  ,\end{equation} where $D(a_{\rho}(f))$ is the domain of $a_{\rho
  }(f)$.  Now let $\{ f_{k}:k=1,\ldots,m \}$ be some $J$-orthonormal basis 
  for $F$, and define
\begin{equation} \label{eq:right} [n_{\rho}(F)](\psi ):=\sum _{k=1}^{m}\norm{ a_{\rho
      }(f_{k})\psi }^{2} ,\end{equation} 
for any $\psi \in D(n_{\rho}(F))$.  It can be shown that the sum in (\ref{eq:right}) 
is independent of the chosen orthonormal basis for $F$, and 
that $D(n_{\rho}(F))$ lies dense in $\hil{H}_{\rho}$.   
Given any densely defined, positive, 
closed quadratic form $t$ on $\hil{H}_{\rho}$, there exists a unique 
positive self-adjoint operator $T$ on $\hil{H}_{\rho}$
such that $D(t)=D(T^{1/2})$ and \begin{equation} 
t(\psi )=\langle T^{1/2}\psi ,T^{1/2}\psi
\rangle , \qquad \psi \in D(t).\end{equation}
We let $N_{\rho}(F)$ denote the finite-subspace $J$-quanta number operator 
on $\hil{H}_{\rho}$ arising from the quadratic 
form $n_{\rho}(F)$.  

We seek a representation-\emph{independent} value for
 ``$\mathrm{Prob}^{\rho}(N(F)\in \Delta)$'', where $\Delta\subseteq \mathbb{N}$. 
 So let $\tau$ be \emph{any} regular state of $\alg{W}$, and let
$N_{\tau}(F)$ be the corresponding number operator on
$\hil{H}_{\tau}$.  Let $\alg{W}_{F}$ be the Weyl algebra over
$(F,\sigma |_{F})$, and let $E_{\tau }(F)$ denote the spectral measure for
$N_{\tau}(F)$ acting on $\hil{H}_{\tau}$.  Then,
$[E_{\tau}(F)](\Delta )$ (the spectral projection representing the 
proposition ``$N_{\tau}(F)\in \Delta$'') is in the weak closure of $\pi _{\tau}
(\alg{W}_{F})$, by the
Stone-von Neumann uniqueness theorem. In particular, there is a net $\{
A_{i}\} \subseteq \alg{W}_{F}$ such that $\pi _{\tau} (A_{i})$
converges weakly to $[E_{\tau}(F)](\Delta )$.  Now, the Stone-von
Neumann uniqueness theorem also entails that
there is a density operator $D_{\rho}$ on $\hil{H}_{\tau}$ such that
\begin{equation} \rho (A)=\mathrm{Tr}(D_{\rho}\pi _{\tau} (A)) ,\qquad A\in
  \alg{W}_{F} .\end{equation}  
We therefore define
\begin{eqnarray} \mathrm{Prob}^{\rho}(N(F)\in \Delta) &:=& \lim
  _{i}\rho (A_{i}) \label{abstract} \\
&=& \lim _{i}\mathrm{Tr}(D_{\rho}\pi _{\tau} (A_{i})) \\
&=&\mathrm{Tr}(D_{\rho}[E_{\tau}(F)](\Delta )) .\label{finite} \end{eqnarray}
The final equality displays that this definition is independent of the chosen
  approximating net $\{ \pi _{\tau} (A_{i}) \}$, and the penultimate
  equality displays that this definition is independent of the (regular)
  representation $\pi _{\tau}$.   In particular, since we may take
$\tau =\rho$,
  it follows that 
  \begin{eqnarray}
\mathrm{Prob}^{\rho}(N(F)\in \Delta)&=&\langle \Omega
_{\rho},[E_{\rho}(F)](\Delta )\Omega _{\rho}\rangle ,\end{eqnarray}
exactly as expected.

We can also define a positive, closed quadratic form on $\hil{H}_{\rho}$
corresponding to the \emph{total} $J$-quanta number operator by:
\begin{eqnarray}
n_{\rho}(\psi )=\sup _{F\in \fin}\,[n_{\rho}(F)](\psi ) \, ,\qquad \qquad \\
D(n_{\rho})=\left\{ \psi\in\hil{H}_{\rho}:\psi \in \bigcap _{f\in
    S}D(a_{\rho}(f)),\, n_{\rho}(\psi )<\infty \right\}
,\end{eqnarray}
where $\fin$ denotes the collection of all finite-dimensional 
subspaces of $\hil{S}_{J}$.  If $D(n_{\rho})$ is dense in $\hil{H}_{\rho}$, then it makes sense to
say that the total $J$-quanta number operator $N_{\rho}$ exists on the
Hilbert space $\hil{H}_{\rho}$.  In general, however, $D(n_{\rho})$ will not be
dense, and may contain only the $0$ vector.  Accordingly, we cannot use a direct
analogue to Eqn.~(\ref{finite}) to define the probability, in the state
$\rho$, that there are, say, $n$ or fewer $J$-quanta.  

However, we can still proceed as 
follows.
Fix $n\in \mathbb{N}$, and suppose $F\subseteq F'$ with both $F,F'\in\fin$.  Since any 
state with $n$ or fewer $J$-quanta with wavefunctions in $F'$ cannot 
have \emph{more} than $n$ $J$-quanta with wavefunctions in the (smaller) subspace 
$F$, 
\begin{equation} \mathrm{Prob}^{\rho}(N(F)\in [0,n]) \:\geq \:
\mathrm{Prob}^{\rho}(N(F')\in [0,n]) .\end{equation}  
Thus, whatever value we obtain for 
``$\mathrm{Prob}^{\rho}(N\in [0,n])$'', it should satisfy the
inequality \begin{equation} 
\mathrm{Prob}^{\rho}(N(F)\in [0,n]) \: \geq \: 
\mathrm{Prob}^{\rho}(N\in [0,n]), \end{equation}
 for any
finite-dimensional subspace $F\subseteq\hil{S}_{J}$.  However, the 
following result holds.

\begin{prop} If $\rho$ is a regular state of $\alg{W}$ disjoint from 
the Fock state $\omega _{J}$, then $\inf _{F\in \fin} \, \Bigl\{ \,
    \mathrm{Prob}\,^{\rho}(N_{F}\in [0,n]) \Bigr\} =0$ for every $n\in \mathbb{N}$.
    \label{chaiken} 
\end{prop} 
\noindent Thus $\rho$ must assign every finite number of 
$J$-quanta probability zero; i.e., $\rho$ predicts an infinite number of 
$J$-quanta with probability 1!
  
Let us tighten this up some more.  Suppose that we are in any regular representation $(\pi
_{\omega},\hil{H}_{\omega})$ in which the total $J$-quanta number operator
$N_{\omega}$ exists and is affiliated to $\pi
_{\omega}(\alg{W})''$.  (For example, we may take the Fock
representation where $\omega =\omega _{J}$.)  Let $E_{\omega}$ denote the
spectral measure of $N_{\omega}$ on
$\hil{H}_{\omega}$.  Considering $\rho$ as a
state of $\pi _{\omega }(\alg{W})$, it is then reasonable to define
\begin{equation} \label{eq:defn} 
\mathrm{Prob}^{\rho}(N\in [0,n]):=\hat{\rho}(E_{\omega}([0,n])),\end{equation} 
where $\hat{\rho}$ is any extension of $\rho$ 
to
$\pi _{\omega }(\alg{W})''$, provided the right-hand side takes the 
same value for all extensions.  (And, of course, it will when 
$\rho\in\mathfrak{F}(\pi _{\omega})$, where (\ref{eq:defn}) reduces to the standard 
definition.)  Now clearly
\begin{equation}
[E_{\omega}(F)]([0,n]) \: \geq \: E_{\omega}([0,n]) ,\ F\in\fin.\end{equation}
(``If there are at most $n$ $J$-quanta in total, then there are at 
most $n$ $J$-quanta whose wavefunctions lie in any finite-dimensional 
subspace of $\hil{S}_{J}$''.)
Since states preserve order relations between projections, every 
extension $\hat{\rho}$ must therefore satisfy 
\begin{equation}
\mathrm{Prob}^{\rho}(N(F)\in [0,n]) \:=\:\hat{\rho}([E_{\omega}(F)]([0,n]))
\: \geq \: \hat{\rho}(E_{\omega}([0,n])) .\end{equation}
Thus, if $\rho$ is disjoint from $\omega$, Proposition~\ref{chaiken} entails that
$\mathrm{Prob}^{\rho}(N\in [0,n])=0$ for all finite $n$.\footnote{Notice that such a prediction could never be made by a state in the 
folium of $\pi
_{\omega}$, since normal states are countably additive (see note 
\ref{count}).}  

As an immediate consequence of this  and the 
disjointness of the Minkowski and Rindler representations, we have 
(reverting back to our earlier number operator notation): 
\begin{equation}  \label{eq:goforit}
\mathrm{Prob}^{\omega_{M}^{\bowtie}}(N_{R}\in [0,n])=0
=\mathrm{Prob}^{\omega_{R}^{\bowtie}}(N_{M}\in [0,n]),\ \mbox{for 
all}\ n\in \mathbb{N},
\end{equation}
\begin{equation} \label{eq:fart}
\mathrm{Prob}^{\omega_{M}^{\triangleright}}(N_{\triangleright}\in [0,n])=0
=\mathrm{Prob}^{\omega_{M}^{\triangleleft}}(N_{\triangleleft}\in [0,n]),\ \mbox{for 
all}\ n\in \mathbb{N}.
\end{equation}
The same probabilities obtain when the Minkowski vacuum is replaced 
with any other state normal in the Minkowski representation.\footnote{This underscores the utter bankruptcy, from the standpoint 
of the liberal about observables, in taking the weak 
equivalence of the Minkowski and Rindler 
representations to be sufficient for their physical 
equivalence.   Yes, every Rindler state of the Weyl algebra is 
a weak* limit of 
Minkowski states.  But the former all predict a finite number of 
Rindler quanta with probability 1, while the latter all predict an 
\emph{infinite} number with probability 1! (Wald ([1994], pp. 82-3) 
makes the exact same point with respect to states 
that do and do not satisfy the ``Hadamard'' property.)} 
So it could not be farther from the truth to say that there is merely the 
potential for Rindler 
quanta in the Minkowski vacuum, or any other eigenstate of $N_{M}$.  

One must be careful, however, with an informal statement like ``The
$M$-vacuum contains infinitely many $R$-quanta with probability 1''.  
Since Rindler wedges are unbounded, there is nothing unphysical, or 
otherwise metaphysically incoherent, about 
thinking of wedges as containing an infinite number of Rindler 
quanta.  But we must not equate this with the quite different \emph{empirical}
claim 
``A Rindler observer's particle detector
has the sure-fire disposition to register the value `$\infty$' ''.   There is no 
such value!  Rather, the empirical content of
 equations (\ref{eq:goforit}) and (\ref{eq:fart}) 
 is simply that an idealized ``two-state'' measuring
apparatus designed to register whether there are $> n$ Rindler quanta
in the Minkowski vacuum will always return the answer `Yes'.  This is
a perfectly sensible physical disposition for a measuring device to
have.  Of course, we are not pretending to have in hand a 
specification of the physical details of such a device.  Indeed, when 
physicists model particle detectors, 
these are usually assumed to 
couple to specific ``modes'' of the field, represented by 
finite-subspace, not total, number operators (cf., e.g., Wald [1994], 
Sec. 3.3).  But this is really beside the 
point, since Teller advertises his resolution of the paradox as a 
way to  \emph{avoid}  a ``retreat to instrumentalism'' about the particle 
concept ([1995], p. 110).  

On Teller's behalf, one might object that there are still no grounds 
for saying any $R$-quanta obtain in the $M$-vacuum, since for any 
particular number $n$ of $R$-quanta you care to name, 
equations (\ref{eq:goforit}) and (\ref{eq:fart}) entail that $n$
is \emph{not} the number of $R$-quanta in the $M$-vacuum.  But 
remember that the same is true for $n=0$, and that, therefore, $n\geq 
1$ $R$-quanta has probability 1!  A further tack might be to deny that 
probability $0$ for $n=0$, or any other $n$, entails 
impossibility or non-actuality of that number of $R$-quanta.  
This would be similar to a common move made in 
response to the lottery paradox, in the hypothetical case 
where there are an infinite number of 
ticket holders.  Since \emph{someone} has to win, each ticket holder must 
still have the potential to win, even though his or her probability of 
winning is zero.  The difficulty with this response is that in the 
Rindler case, we have no independent reason to think that some 
particular finite number of $R$-quanta \emph{has} to be detected at 
all.  
Moreover, if we were to go soft on taking probability $0$ to be 
sufficient for ``not actual'', we should equally deny that probability 1 
is sufficient for ``actual'', and by Teller's lights the paradox would go away at a 
stroke (because there could never be actual Rindler \emph{or} Minkowski quanta
 in \emph{any} field state).

We conclude that Teller's resolution of the paradox of 
observer-dependence of particles fails.  And so be it, since it was 
ill-motivated in the first place. 
We already indicated in the previous subsection that
it should be enough of a resolution to recognize that there are
different kinds of quanta.  We believe the physicists of the field and
detector approaches are correct to bite the bullet hard on this, even
though it means abandoning na\"{i}ve realism about particles
(though not, of course, about detection events).  We turn, next, to
arguing that a coherent story can still be told about
the relationship between the different kinds of particle talk used by
different observers.

\subsection{Incommensurable or Complementary?}

At the beginning of this paper, we reproduced a passage from Jauch's 
amusing Galilean dialogue on the question ``Are Quanta Real?''.  
In that passage, Sagredo is glorying in the prospect that 
complementarity may be applicable even in classical physics; and, more 
generally, to 
solving 
the philosophical problem of the specificity of individual 
events versus the generality of scientific description.  
It is well-known that Bohr himself sought to extend the idea of complementarity 
to all different walks of life, beyond its originally intended 
application in quantum theory.  And even within the confines of quantum theory, it is 
often the case 
that 
when the going gets tough, tough quantum theorists cloak themselves in 
the mystical profundity of complementarity, sometimes just 
to get philosophers off their backs.
  
So it seems with the following notorious comments of a well-known advocate of
the detector approach that have received a predictably cool reception
from philosophers:

\begin{quote}
Bohr taught us that quantum mechanics is an algorithm for computing the
results of measurements. Any discussion about what is a ``real, physical
vacuum'', must therefore be related to the behaviour of real, physical
measuring devices, in this case particle-number detectors. Armed with such
heuristic devices, we may then assert the following. There are quantum
states and there are particle detectors. Quantum field theory enables us to
predict probabilistically how a particular detector will respond to that
state. That is all. That is all there can ever be in physics, because
physics is about the observations and measurements that we can make in the
world. We can't talk meaningfully about whether such-and-such a state
contains particles except in the context of a specified particle detector
measurement. To claim (as some authors occasionally do!) that when a
detector responds (registers particles) in somebody's cherished vacuum state
that the particles concerned are ``fictitious'' or ``quasi-particles'', or that
the detector is being ``misled'' or ``distorted'', is an empty statement (Davies
[1984], p. 69).
\end{quote}
\noindent We shall argue that, cleansed of Davies' purely 
operationalist reading 
of Bohr, 
complementarity \emph{does}, after all, shed light on the relation 
between inequivalent particle concepts.

R\"{u}ger [1989] balks at this idea.  He writes:  
\begin{quote}
  The ``real problem'' --- how to understand how there might be
  particles for one observer, but none at all for another observer in
  a different state of motion --- is not readily solved by an appeal
  to Copenhagenism... Though quantum mechanics can tell us that the
  \textit{properties} of micro-objects (like momentum or energy)
  depend in a sense on observers measuring them, the standard
  interpretation of the theory still does not tell us that whether
  there is a micro-object or not depends on observers. At least the
  common form of this interpretation is not of immediate help here
  (R\"{u}ger [1989], pp. 575-6).
\end{quote}

\noindent Well, let us consider the ``common form'' of the Copenhagen interpretation.  
Whatever one's preferred embellishment of the interpretation, it must
at least imply that observables represented by noncommuting
``complementary'' self-adjoint operators cannot have simultaneously
determinate values in all states.  Since field quantizations are built
upon an abstract noncommutative algebra, the Weyl algebra,
complementarity retains its application to quantum field theory.  In
particular, in any \emph{single} Fock space representation --- setting
aside inequivalent representations for the moment --- there will be a
total number operator and nontrivial superpositions of its
eigenstates. In these superpositions, which are eigenstates of
observables failing to commute with the number operator, it is
therefore perfectly in line with complementarity that we say they
contain no actual particles in any substantive sense.\footnote{As
  R\"{u}ger notes earlier ([1989], p. 571), in ordinary
  non-field-theoretic quantum theory, complementarity only undermined
  a na\"{i}ve substance-properties ontology.  However, this was only
  because there was no ``number of quanta'' observable in the theory!}
In addition, there will be different number operators on Fock space
that count the number of quanta with wavefunctions lying in different
subspaces of the one-particle space, and they will only commute if the
corresponding subspaces are compatible.  So even before we consider
inequivalent particle concepts, we must already accept that there are
different \emph{complementary} ``kinds'' of quanta, according to what
their wavefunctions are.

Does complementarity extend to the particle concepts associated with 
inequivalent Fock representations?  \emph{Contra}
 R\"{u}ger [1989], we claim that it does.  We saw 
earlier that one can build finite-subspace $J$-quanta number operators 
in \emph{any} regular 
representation of $\alg{W}[S,\sigma]$, provided only that $J$ defines a 
proper complex structure on $S$ that leaves it 
invariant.  In particular, using the canonical commutation relation 
$[\Phi(f),\Phi(g)]=i\sigma(f,g)I$, 
a tedious but elementary calculation reveals that, for any $f,g\in S$,
\begin{eqnarray} & [N_{J_{1}}(f),N_{J_{2}}(g)] \nonumber \\ \label{eq:ha!}
 = & i/2\{\sigma(f,g)[\Phi(f),\Phi(g)]_{+}+
\sigma(f,J_{2}g)[\Phi(f),\Phi(J_{2}g)]_{+} \\  \nonumber
& +\sigma(J_{1}f,g)[\Phi(J_{1}f),\Phi(g)]_{+}
+\sigma(J_{1}f,J_{2}g)[\Phi(J_{1}f),\Phi(J_{2}g)]_{+}\},
\end{eqnarray}
in any regular representation.\footnote{As a check on expression
  (\ref{eq:ha!}), note that it is invariant under the one-particle
  space phase transformations $f\rightarrow (\cos t+J_{1}\sin t)f$ and
  $g\rightarrow (\cos t+J_{2}\sin t)g$, and when $J_{1}=J_{2}=J$,
  reduces to zero just in case the rays generated by $f$ and $g$ are
  compatible subspaces of $\hil{S}_{J}$.}  Thus, there are
well-defined and, in general, \emph{nontrivial} commutation relations
between finite-subspace number operators, even when the associated
particle concepts are inequivalent.  We also saw in Eqn.
(\ref{eq:invariant?})  that when $J_{2}\not=J_{1}$, no $N_{J_{2}}(f)$,
for any $f\in \hil{S}_{J_{2}}$, will leave the zero-particle subspace
of $N_{J_{1}}$ invariant.  Since it is a necessary condition that this
nondegenerate eigenspace be left invariant by any self-adjoint
operator commuting with $N_{J_{1}}$, it follows that
$[N_{J_{2}}(f),N_{J_{1}}]\not=0$ for all $f\in \hil{S}_{J_{2}}$.  Thus
finite-subspace number operators for one kind of quanta are
complementary to the total number operators of inequivalent kinds of
quanta.

Of course, we cannot give the same argument for 
complementarity between the \emph{total} number operators 
$N_{J_{1}}$ and $N_{J_{2}}$ pertaining to 
inequivalent kinds of quanta, because, as we know, they cannot even be defined as 
operators on the 
same Hilbert space.  However, we disagree with Arageorgis ([1995], pp. 
303-4) that this means Teller's ``complementarity talk'' in relation 
to the Minkowski and Rindler total number operators is wholly inapplicable.  
We have two reasons for the 
disagreement.  

First, since it is a necessary condition that a (possibly unbounded)
self-adjoint observable $Y$ on $\hil{H}_{\omega_{J_{1}}}$ commuting
with $N_{J_{1}}$ have $\Omega_{\omega_{J_{1}}}$ as an eigenvector, it
is also necessary that the abstract vacuum state $\omega_{J_{1}}$ be
dispersion-free on $Y$.  But this latter condition is purely algebraic
and makes sense even when $Y$ does \emph{not} act on
$\hil{H}_{\omega_{J_{1}}}$.  Moreover, as Proposition \ref{teller}
shows, this condition fails when $Y$ is taken to be the total number
operator of any Fock representation inequivalent to
$\pi_{\omega_{J_{1}}}$.  So it is entirely natural to treat
Proposition \ref{teller} as a vindication of the idea that
inequivalent pairs of total number operators are complementary.

Secondly, we have seen that any state in the folium of a representation 
associated with one kind of quanta 
assigns probability zero to any finite number of an inequivalent kind 
of quanta. This has a direct 
analogue in the most famous instance of complementarity: that which 
obtains between the concepts of position and momentum.  

Consider the 
unbounded position and momentum operators, $x$ and $p$ 
($=-i\frac{\partial}{\partial 
x}$), acting on $L^{2}(\mathbb{R})$.  Let $E_{x}$ and $E_{p}$ be their 
spectral measures.  We say that a state $\rho$ of 
$\mathbf{B}(L^{2}(\mathbb{R}))$ assigns $x$ a \emph{finite} 
dispersion-free value just in case $\rho$ is dispersion-free on $x$ and 
there is a $\lambda\in \mathbb{R}$ such that $\rho(E_{x}((a,b)))=1$ if and 
only if $\lambda\in (a,b)$.  (Similarly, for $p$.)   Then the following 
is a direct 
consequence of the canonical commutation relation 
$[x,p]=iI$ (see Halvorson and Clifton [1999], Prop. 3.7).
\begin{prop}  \label{us} If $\rho$ is a state of 
$\mathbf{B}(L^{2}(\mathbb{R}))$ that assigns $x$ (resp., $p$) a finite 
dispersion-free value, then $\rho(E_{p}((a,b)))=0$ (resp., 
$\rho(E_{x}((a,b)))=0$) for any $a,b\in\mathbb{R}$.  
\end{prop}
\noindent This result makes rigorous the fact, 
suggested by Fourier analysis, that if either of $x$ or $p$ has a 
sharp finite value in any state, the other is ``maximally indeterminate''. 
But the same goes for pairs of inequivalent number operators 
$(N_{J_{1}},N_{J_{2}})$: if a regular state $\rho$ assigns $N_{J_{1}}$ a 
finite dispersion-free value, then 
$\rho\in\mathfrak{F}(\pi_{\omega_{J_{1}}})$ which, in turn, entails 
that $\rho$ assigns probability zero to any finite set of eigenvalues 
for $N_{J_{2}}$.  Thus, $(N_{J_{1}},N_{J_{2}})$ 
are, in a natural sense, \emph{maximally} complementary, despite the fact 
that they have no well-defined commutator.

One might object that our analogy is only skin deep; after all, $x$
and $p$ still act on the \emph{same} Hilbert space,
$L^{2}(\mathbb{R})$!  So let us deepen the analogy.  Let $\alg{W}$ be
the Weyl algebra for one degree of freedom, and let $U(a)\equiv
W(a,0)$ and $V(b)\equiv W(0,b)$ be the unitary operators
corresponding, respectively, to position and momentum.  Now, if we
think of position as analogous to the Minkowski number operator and
momentum as analogous to the Rindler number operator, the standard
Schr{\"o}dinger representation is not the analogue of the Minkowski
vacuum representation --- since the Minkowski vacuum representation is
constructed so as to have eigenvectors for $N_{M}$, whereas the
Schr{\"o}dinger representation obviously does not have eigenvectors
for $x$.  Thus, to find a representation analogous to the Minkowski
vacuum representation, first choose a state $\rho$ of $\alg{W}$ that is
dispersion-free on all elements $\{ U(a):a\in \mathbb{R} \}$.  In
particular, we may choose $\rho$ such that $\rho (U(a))=e^{ia\lambda
  }$ for all $a \in \mathbb{R}$.  If we then let
$(\pi_{\rho},\hil{H}_{\rho},\Omega_{\rho})$ denote the GNS
representation of $\alg{W}$ induced by $\rho$, it follows that we may
construct an unbounded position operator $x$ on $\hil{H}_{\rho}$ which
has $\Omega_{\rho}$ as an eigenvector with eigenvalue $\lambda$.  But,
lo and behold, it is not possible to define a momentum operator $p$ on
the Hilbert space $\hil{H}_{\rho}$.

Indeed, since $\rho$ is dispersion-free on $U(a)$, it is
multiplicative for the product of $U(a)$ with any other element of
$\alg{W}$ (KR [1997], Ex. 4.6.16).  In particular,
\begin{equation} 
\rho(U(a))\rho(V(b))=
  e^{iab}\rho(V(b))\rho(U(a)),\qquad a,b\in \mathbb{R}.
\end{equation}
Since $\rho(U(a))=e^{ia\lambda}\not=0$, this implies
\begin{equation} 
\rho(V(b))=e^{iab}\rho(V(b)),\qquad a,b\in \mathbb{R}. \label{eq:cannot}
\end{equation}
However, when $a\not=0$, (\ref{eq:cannot}) cannot hold for all
$b\not=0$ unless $\rho(V(b))=0$.  Thus, 
\begin{equation}
\langle\Omega_{\rho},\pi_{\rho}(V(b)) \Omega_{\rho}\rangle=0,\qquad \forall b\not=0.
\end{equation}
On the other hand,
\begin{equation}
\langle\Omega_{\rho},\pi_{\rho}(V(0))\Omega_{\rho}\rangle 
=\langle\Omega_{\rho}, I\Omega_{\rho}\rangle=1.
\end{equation}
Thus, $\pi_{\rho}(V(b))$ is not weakly continuous in $b$, 
and there can be no self-adjoint operator $p$ on 
$\hil{H}_{\rho}$ such that $V(b)=e^{ibp}$.   On the other hand, 
since $a\in \mathbb{R}\mapsto \rho(U(a))=e^{ia\lambda}$ is continuous, and 
hence $\pi_{\rho}$ is regular with respect to the subgroup of unitary 
operators $\{U(a):a\in \mathbb{R}\}$, there \emph{is} a position operator 
on $\hil{H}_{\rho}$.   

Similarly, if $\omega$ is a state of $\alg{W}$ that is dispersion-free
on the momentum unitary operators $\{ V(b):b\in \mathbb{R} \}$, then
it is not possible to define a position operator on the Hilbert space
$\hil{H}_{\omega}$.  Moreover, the GNS representations $\pi _{\rho}$
and $\pi _{\omega}$ are disjoint --- precisely as in the case of the
GNS representations induced by the Minkowski and Rindler vacuum
states.  Indeed, suppose for reductio that there is a unitary operator
$T$ from $\hil{H}_{\omega}$ to $\hil{H}_{\rho}$ such that $T^{-1}\pi
_{\rho}(A)T=\pi _{\omega }(A)$ for all $A\in \alg{W}$.  Then, it would
follow that $\pi _{\omega }(U(a))=T^{-1}\pi _{\rho }(U(a))T$ is weakly
continuous in $a$, in contradiction to the fact that $x$ cannot be
defined on $\hil{H}_{\omega}$.
 
So we maintain that there are compelling formal reasons for thinking of
Minkowski and Rindler quanta as complementary.  What's more, when a
Minkowski observer sets out to detect particles, her state of motion
determines that her detector will be sensitive to the presence of
Minkowski quanta.  Similarly for a Rindler observer and his detector.
This is borne out by the analysis of Unruh and Wald [1984] in which
they show how his detector will \emph{itself} ``define'' (in a
``nonstandard'' way) what solutions of the relativistic wave equation
are counted as having positive frequency, via the way the detector
couples to the field. So we may think of the choice of an observer to
follow an inertial or Rindler trajectory through spacetime as
analogous to the choice between measuring the position or momentum of
a particle.  Each choice requires a distinct kind of coupling to the
system, and both measurements cannot be executed on the field
simultaneously and with infinite precision.\footnote{Why can't
  \emph{both} a Minkowski and a Rindler observer set off in different
  spacetime directions and \emph{simultaneously} measure their
  respective (finite-subspace or total) number operators?  Would it
  not, then, be a violation of microcausality when the Minkowski
  observer's measurement disturbs the statistics of the Rindler
  observer's measurement outcomes?  No.  We must remember that the
  Minkowski particle concept is global, so our Minkowski observer
  cannot make a precise measurement of any of her number operators
  unless it is executed throughout the whole of spacetime, which would
  necessarily destroy her spacelike separation from the Rindler
  observer.  On the other hand, if she is content with only an
  approximate measurement of one of her number operators in a bounded
  spacetime region, it is well-known that simultaneous, nondisturbing
  ``unsharp'' measurements of incompatible observables \emph{are}
  possible.  For an analysis of the case of simultaneous measurements
  of unsharp position and momentum, see Busch \emph{et al} [1995].}
Moreover, execution of one type of measurement precludes meaningful
discourse about the values of the observable that the observer did not
choose to measure. All this is the essence of ``Copenhagenism''.
 
 And it should \emph{not} be equated with operationalism!  The goal of
 the detector approach to the paradox of observer-dependence was to
 achieve clarity on the problem by reverting back to operational
 definitions of the word ``particle'' with respect to the concrete
 behaviour of particular kinds of detectors (cf., e.g., DeWitt
 [1979b], p. 692).  But, as with early days of special relativity and
 quantum theory, operationalism can serve its purpose and then be
 jettisoned.  Rindler quanta get their status as such not because they
 are, \emph{by definition}, the sort of thing that accelerated
 detectors detect.  This gets things backwards. Rindler detectors
 display Rindler quanta in the Minkowski vacuum \emph{because} they
 couple to \emph{Rindler} observables of the field that are distinct
 from, and indeed complementary to, Minkowski observables.
 
 Arageorgis [1995]  himself, together with his collaborators 
(Arageorgis \emph{et al} [1995]), prefer to
 characterize inequivalent particle 
concepts, not as complementary, but \emph{incommensurable}.  At first 
glance, this looks like a trivial semantic dispute between us.  
For instance Glymour, in a recent introductory 
text on the philosophy of science, summarizes complementarity 
using the language of incommensurability: 
\begin{quote}
Changing the
experiments we conduct is like changing conceptual schemes or paradigms: we
experience a different world. Just as no world of experience combines
different conceptual schemes, no reality we can experience (even indirectly
through our experiments) combines precise position and precise 
momentum (Salmon \emph{et al} [1992], p.
128). \end{quote}
However,
philosophers of science usually think of incommensurability as a 
relation between theories \emph{in toto}, not different parts of the same 
physical theory.  Arageorgis \emph{et al} maintain that inequivalent 
quantizations define incommensurable \emph{theories}.     

Arageorgis [1995] makes the claim that ``the degrees of freedom of the field in the Rindler model
\emph{simply cannot be described} in terms of the ground state and the elementary
excitations of the degrees of freedom of the field in the Minkowski 
model'' ([1995], p. 268; our italics).  Yet so much of our earlier 
discussion proves the contrary.  Disjoint representations 
\emph{are} commensurable, via the abstract Weyl algebra they share.  
The result is that the ground state of one Fock representation makes
 definite, if sometimes counterintuitive, predictions for the
  ``differently complexified'' degrees of freedom of other Fock 
 representations.
 
 Arageorgis \emph{et al} 
 [2000] offer an \emph{argument}
  for incommensurability --- based on Fulling's ``theorem''.  They 
 begin by discussing the case where the primed and unprimed 
 representations are unitarily equivalent.  (Notice that they speak of 
 two different ``theorists'', rather than two different observers.)   
 
 \begin{quote}
...while \textit{different}, these particle concepts can nevertheless be
deemed to be \textit{commensurable}. The two theorists are just labelling
the particle states in different ways, since each defines particles of a
given type by mixing the creation and annihilation operators of the other
theorist. Insofar as the primed and unprimed theorists disagree, they
disagree over which of two inter-translatable descriptions of the same
physical situation to use.\smallskip 

The gulf of disagreement between two theorists using unitarily
inequivalent Fock space representations is much deeper. If in this
case the primed-particle theorist can speak sensibly of the
unprimed-particle theorist's vacuum at all, he will say that its
primed-particle content is infinite (or more properly, undefined), and
the unprimed-theorist will say the same of the unprimed-particle
content of the primed vacuum. Such disagreement is profound enough
that we deem the particle concepts affiliated with unitarily
inequivalent Fock representations \textit{incommensurable} ([2000], p.
26).
\end{quote}

The logic of this argument is curious.  
In order to make Fulling's 
``theorem'' do the work for incommensurability that Arageorgis \emph{et 
al} want it to, one must first have in hand a rigorous version of the 
theorem (otherwise their argument would be built on sand).  
But any rigorous version, like our Proposition \ref{fulling}, has to presuppose
that there is sense to be made of using a vector state from one Fock 
representation to generate a prediction for the expectation value of 
the total number operator in another inequivalent representation.  
Thus, one cannot even \emph{entertain}
 the philosophical implications of Fulling's result if one 
has not first 
granted a certain level of commensurability between inequivalent representations.  

Moreover, while it may be tempting to \emph{define} what one means by
``incommensurable representations'' in terms of Fulling's
characterization of inequivalent representations, it is difficult to
see the exact motivation for such a definition.  Even vector states
\emph{in the folium} of the unprimed ``theorist's'' Fock
representation can fail to assign his total number operator a finite
expectation value (just consider any vector not in the operator's
domain). Yet it would be alarmist to claim that, were the field in
such a state, the unprimed ``theorist'' would lose his conceptual
grasp on, or his ability to talk about, his \emph{own} unprimed kind
of quanta!  So long as a state prescribes a well-defined probability
measure over the spectral projections of the unprimed ``theorist's''
total number operator --- and all states in his \emph{and} the folium
of any primed theorist's representation \emph{will} --- we fail to see
the difficulty.

\section{Conclusion}
  
Let us return to answer the questions we raised in our introduction.
  
We have argued that a conservative operationalist about physical
observables is not committed to the physical inequivalence of disjoint
representations, so long as he has no particular attachment to states
in a particular folium being the only physical ones.  On the other
hand, a liberal about physical observables, no matter what his view on
states, \emph{must} say that disjoint representations yield physically
inequivalent descriptions of a field.  However, we steadfastly
resisted the idea that this means an interpreter of quantum field
theory must say disjoint representations are incommensurable, or even
different, \emph{theories}.

Distinguishing ``potential'' from ``actual'' quanta won't do to
resolve the paradox of observer-dependence.  Rather, the paradox
forces us to thoroughly abandon the idea that Minkowski and Rindler
observers moving through the same field are both trying to detect the
presence of particles \emph{simpliciter}.  Their motions cause their
detectors to couple to \emph{different} incompatible particle
observables of the field, making their perspectives on the field
necessarily complementary.  Furthermore, taking this complementary
seriously means saying that neither the Minkowski nor Rindler
perspective yields the uniquely ``correct'' story about the particle
content of the field, and that \emph{both} are necessary to provide a
complete picture.
 
So ``Are Rindler Quanta Real?''  This is a loaded question that can be
understood in two different ways.
  
First, we could be asking ``Are \emph{Any} Quanta Real?'' without
regard to inequivalent notions of quanta.  Certainly particle
detection events, modulo a resolution of the measurement problem, are
real.  But it should be obvious by now that detection events do not
generally license na\"{i}ve talk of individuatable, localizable,
particles that come in determinate numbers in the \emph{absence} of
being detected.
  
A fuller response would be that quantum field theory is
``fundamentally'' a theory of a field, not particles.  This is a
reasonable response given that: (i) the field operators
$\{\Phi(f):f\in S\}$ exist in every regular representation; (ii) they
can be used to construct creation, annihilation, and number operators;
and (iii) their expectation values evolve in significant respects like
the values of the counterpart classical field, modulo non-local
Bell-type correlations.  This ``field approach'' response might seem
to leave the ontology of the theory somewhat opaque.  The field
operators, being subject to the canonical commutation relations, do
not all commute; so we cannot speak sensibly of them all
simultaneously having determinate values!  However, the right way to
think of the field approach, compatible with complementary, is to see
it as viewing a quantum field as a collection of correlated
``objective propensities'' to display values of the field operators in
more or less localized regions of spacetime, relative to various
measurement contexts.  This view makes room for the reality of quanta,
but only as a kind of epiphenomenon of the field associated with
certain functions of the field operators.
  
  Second, we could be specifically interested in knowing whether it is 
  sensible to say that \emph{Rindler}, as opposed to just 
  Minkowski, quanta are real.  An uninteresting answer would be 
  `No' --- on the grounds that quantum field theory on flat spacetime 
  is not a serious candidate for describing our actual universe, or 
  that the Rindler representation is too ``pathological''.  But, as 
  philosophers, we are content to leave to the physicists the task of 
  deciding 
  the question ``Are Rindler 
  Quanta 
  \emph{Empirically Verified}?''.  
  All we have tried to determine (to echo words of van Fraassen)
  is how the world \emph{could possibly be} if both the Rindler and Minkowski 
  representations were ``true''.  We have argued that the antecedent 
  of this counterfactual makes perfect sense, and that it forces us to 
  view Rindler and Minkowski quanta as complementary.  Thus, 
  Rindler and Minkowski 
  would be equally amenable to achieving ``reality status'' provided the appropriate 
  measurement context is in place.  
  As Wald has put it:
  \begin{quote}
    Rindler particles are ``real'' to accelerating observers!  This
    shows that different notions of ``particle'' are useful for
    different purposes ([1994], p. 116).
\end{quote}
  \vspace{1em}

\noindent\emph{Acknowledgements:} We are extremely
grateful to John Earman for many stimulating discussions which
provided the impetus for writing this paper, and to Aristidis
Arageorgis for writing a provocative and inspiring dissertation.  We
would also like to thank Klaas Landsman and Rainer Verch for help with
the proof of Proposition~\ref{segal}. \vspace{1em}

{\flushright \emph{Department of Philosophy} \\
  \emph{University of Pittsburgh} \\
  \emph{Pittsburgh, PA 15260}  \\
  \emph{(e-mails: rclifton@pitt.edu, hphst1@pitt.edu)} \\ }
\vspace{2em}

\section*{Appendix}
\noindent \textbf{Proposition \ref{frog}}. \emph{Under the liberal approach to observables, 
  $\phi$ (factorial) and $\pi$ (irreducible) are physically equivalent
  representations of $\alg{W}$ only if they are quasi-equivalent.}

\begin{proof} By hypothesis, the bijective mapping 
  $\alpha$ must map the self-adjoint part of $\phi(\alg{W})''$ onto
  that of $\pi(\alg{W})''$.  Extend $\alpha$ to \emph{all} of
  $\phi(\alg{W})''$ by defining
\begin{equation}
\alpha(X):= 
\alpha(\mathrm{Re}(X))+i\alpha(\mathrm{Im}(X)),\qquad X\in\phi(\alg{W})''.  
\end{equation}
Clearly, then, $\alpha$ preserves adjoints.   

Recall that a family of states $S_{0}$ on a $C^{*}$-algebra is called
\emph{full} just in case $S_{0}$ is convex, and for any $A\in
\alg{A}$, $\rho (A)\geq 0$ for all $\rho \in S_{0}$ only if $A\geq 0$.
By hypothesis, there is a bijective mapping $\beta$ from the
``physical'' states of $\phi (\alg{W})''$ onto the ``physical'' states
of $\pi (\alg{W})''$.  According to both the conservative and liberal
construals of physical states, the set of physical states includes
normal states.  Since the normal states are full, the domain and range
of $\beta$ contain full sets of states of the respective
$C^{*}$-algebras.

By condition (\ref{eq:y}) and the fact that the domain and range of
$\beta$ are full sets of states, $\alpha$ arises from a
\emph{symmetry} between the $C^{*}$-algebras $\phi(\alg{W})''$ and
$\pi(\alg{W})''$ in the sense of Roberts \& Roepstorff ([1969], Sec.
3).\footnote{Actually, they consider only symmetries of a
  $C^{*}$-algebra onto \emph{itself}, but their results remain valid
  for our case.}  Their Propositions 3.1 and 6.3 then apply to
guarantee that $\alpha$ must be linear and preserve Jordan structure
(i.e., anti-commutator brackets).  Thus $\alpha$ is a Jordan
$*$-isomorphism.

Now both $\phi(\alg{W})''$ and
$\pi(\alg{W})''=\mathbf{B}(\hil{H}_{\pi})$ are von Neumann algebras,
and the latter has a trivial commutant.  Thus KR ([1997], Ex. 10.5.26)
applies, and $\alpha$ is either a $*$-isomorphism or a
$*$-anti-isomorphism, that reverses the order of products.  However,
such reversal is ruled out, otherwise we would have, using the Weyl
relations (\ref{eq:herro}),
\begin{eqnarray}
& \alpha(\phi(W(f))\phi(W(g))) = e^{-i\sigma (f,g)/2}\alpha(\phi(W(f+g))),\\
\Rightarrow & \alpha(\phi(W(g)))\alpha(\phi(W(f))) =
e^{-i\sigma (f,g)/2}\alpha(\phi(W(f+g))), \\
\Rightarrow & \pi(W(g))\pi(W(f)) = 
e^{-i\sigma (f,g)/2}\pi(W(f+g)), \\ 
\Rightarrow & e^{i\sigma (f,g)/2}\pi(W(f+g)) = 
e^{-i\sigma (f,g)/2}\pi(W(f+g)), 
\end{eqnarray} 
for all $f,g\in S$.  This entails that the value of $\sigma$ on any 
pair of vectors is always is a multiple of 
$2\pi$ which, since $\sigma$ is bilinear, cannot 
happen unless $\sigma=0$ identically (and hence $S=\{0\}$). It follows that $\alpha$ 
is in fact a $*$-isomorphism.  And, by condition (\ref{eq:x}), 
$\alpha$ must map $\phi(A)$ to $\pi(A)$ for all $A\in\alg{W}$.  Thus 
$\phi$ is quasi-equivalent to $\pi$. \end{proof}

\noindent \textbf{Proposition \ref{segal}.} \emph{When $S$ is 
infinite-dimensional, $\pi(\alg{W}[S,\sigma])$
contains no non-trivial bounded functions of the total number operator
  on $\hil{F}(\hil{S}_{J})$.}  

\begin{proof} 
For clarity, we suppress the representation map $\pi$.  Suppose that 
$F:\mathbb{N}\mapsto \mathbb{C}$ is a bounded function.
  We show that if $F(N)\in \alg{W}$, then $F(n)=F(n+1)$ for all $n\in
  \mathbb{N}$.
  
  The Weyl operators on $\hil{F}(\hil{S}_{J})$ satisfy the commutation 
  relation (BR [1996], Prop. 5.2.4(1,2)):
  \begin{equation} \label{eq:note} W(g)\Phi (f)W(g)^{*}=\Phi (f)-\sigma(g,f)I
    .\end{equation} Using Eqns. (\ref{eq:ip}) and (\ref{eq:two}), we
  find \begin{equation} W(g)a^{*}(f)W(g)^{*}=a^{*}(f)+2^{-1/2}i(g,f)_{J}I
    ,\end{equation} and from this, $[W(g),a^{*}(f)]=2^{-1/2}i(g,f)_{J}W(g)$.  Now
  let $\psi \in \hil{F}(\hil{S}_{J})$ be in the domain of
  $a^{*}(f)$.  Then a straightforward calculation shows that
  \begin{eqnarray} \lefteqn{ \langle a^{*}(f)\psi ,W(g)a^{*}(f)\psi
  \rangle } \qquad \nonumber \\
&=& 2^{-1/2}i(g,f)_{J}\langle a^{*}(f)\psi ,W(g)\psi \rangle +\langle a(f)a^{*}(f)\psi
    ,W(g)\psi \rangle .\label{eq:landsman} \end{eqnarray}
  
  Let $\{ f_{k}\}$ be an infinite orthonormal basis for $\hil{S}_{J}$, and let
  $\psi \in \hil{F}(\hil{S}_{J})$ be the vector whose $n$-th component is
  $P_{+}(f_{1}\otimes \cdots \otimes f_{n})$ and whose other
  components are zero.  Now, for any $k>n$, we have
  $a(f_{k})a^{*}(f_{k})\psi =(n+1)\psi$.  Thus, Eqn. (\ref{eq:landsman})
  gives \begin{eqnarray} \lefteqn{\langle a^{*}(f_{k})\psi
      ,W(g)a^{*}(f_{k})\psi \rangle }
    \qquad \nonumber \\
    &=& 2^{-1/2}i(g,f_{k})_{J}\langle a^{*}(f_{k})\psi ,W(g)\psi \rangle
    +(n+1)\langle \psi ,W(g)\psi \rangle .\end{eqnarray} Hence,
  \begin{equation}
\lim _{k\rightarrow\infty} \,\langle a^{*}(f_{k})\psi ,W(g)a^{*}(f_{k})\psi \rangle =
(n+1)\langle \psi ,W(g)\psi \rangle . \label{eq:last} \end{equation}
Since $\alg{W}$ is generated by the $W(g)$, Eqn. (\ref{eq:last}) holds when
$W(g)$ is replaced with any element in $\alg{W}$.  On the other hand, $\psi$ is an
eigenvector with eigenvalue $n$ for $N$ while $a^{*}(f_{k})\psi$ is an
eigenvector with eigenvalue $n+1$ for $N$.  Thus, $\langle \psi
,F(N)\psi \rangle =F(n)\norm{\psi}^{2}$ while
\begin{eqnarray}
\langle a^{*}(f_{k})\psi , F(N)a^{*}(f_{k})\psi \rangle &=&
F(n+1)\,\norm{a^{*}(f_{k})\psi }^{2} \\
&=& (n+1)F(n+1)\,\norm{\psi}^{2} ,\end{eqnarray}
for all $k>n$.  Thus, the assumption that $F(N)$ is in $\alg{W}$ (and 
hence satisfies (\ref{eq:last})) 
entails that $F(n+1)=F(n)$.  \end{proof}

\noindent \textbf{Proposition \ref{disjoint1}.} \emph{The Minkowski and 
Rindler representations of 
  $\alg{W}_{\triangleleft}$ are disjoint.}

\begin{proof}   

By Horuzhy ([1988], Thm. 3.3.4), 
  $\pi _{\omega _{M}^{\triangleleft}}(\alg{W}_{\triangleleft})''$ is 
  a ``type III'' von Neumann algebra which, in particular, contains no 
  atomic projections.  
   Since $\pi _{\omega _{R}^{\triangleleft}}$ is irreducible and 
  $\pi _{\omega _{M}^{\triangleleft}}$ factorial, either 
  $\pi _{\omega _{R}^{\triangleleft}}$ and $\pi _{\omega _{M}^{\triangleleft}}$ are 
  disjoint, or they are quasi-equivalent.  However, since 
  $\pi _{\omega _{R}^{\triangleleft}}(\alg{W}_{\triangleleft})''=
  \mathbf{B}(\hil{F}(\hil{S}(\triangleleft)_{R}))$, the weak closure 
  of the Rindler representation clearly contains atomic 
  projections.   Moreover, $*$-isomorphisms preserve the ordering of projection operators. 
    Thus there can be no $*$-isomorphism of 
  $\pi _{\omega _{M}^{\triangleleft}}(\alg{W}_{\triangleleft})''$ 
  onto $\pi _{\omega _{R}^{\triangleleft}}(\alg{W}_{\triangleleft})''$,  
  and the Minkowski and Rindler representations of $\alg{W}_{\triangleleft}$  
  are disjoint. \end{proof}
  
 \noindent \textbf{Proposition \ref{disjoint2}.} 
\emph{The Minkowski and Rindler representations of
  $\alg{W}_{\bowtie}$ are disjoint.}  

\begin{proof}  Again, we use the fact that  $\pi
  _{\omega _{M}^{\bowtie}}(\alg{W}_{\triangleleft})''$ 
  ($\equiv\pi _{\omega 
  _{M}^{\triangleleft}}(\alg{W}_{\triangleleft})''$) does not 
  contain atomic projections, whereas 
  $\pi_{\omega _{R}^{\bowtie}}(\alg{W}_{\triangleleft})''$ 
  ($\equiv\pi_{\omega _{R}^{\triangleleft}}(\alg{W}_{\triangleleft})''$) does.  
  Suppose, for reductio ad absurdum, that 
$\omega_{R}^{\bowtie}$ and $\omega _{M}^{\bowtie}$ are not disjoint.  
Since both these states are pure, they induce irreducible
representations, which therefore must be unitarily equivalent.  
Thus, there is a \emph{weakly 
continuous} $*$-isomorphism
  $\alpha$ from $\pi _{\omega _{M}^{\bowtie}}(\alg{W}_{\bowtie})''$ onto $\pi
  _{\omega_{R}^{\bowtie}}(\alg{W}_{\bowtie})''$ such 
that $\alpha (\pi _{\omega _{M}^{\bowtie}}(A))=\pi
  _{\omega_{R}^{\bowtie}}(A)$ for each $A\in 
\alg{W}_{\bowtie}$.  In particular,
  $\alpha$ maps $\pi _{\omega _{M}^{\bowtie}}(\alg{W}_{\triangleleft})$ onto 
  $\pi_{\omega_{R}^{\bowtie}}
  (\alg{W}_{\triangleleft})$; and, 
since $\alpha$ is weakly continuous, it
  maps $\pi _{\omega _{M}^{\bowtie}}(\alg{W}_{\triangleleft})''$ onto 
  $\pi_{\omega_{R}^{\bowtie}}(\alg{W}_{\triangleleft})''$.  Consequently, $\pi
  _{\omega _{M}^{\bowtie}}(\alg{W}_{\triangleleft})''$ contains an atomic projection --- 
  contradiction.  \end{proof} 
  
  \noindent \textbf{Proposition \ref{teller}.} \emph{If $J_{1},J_{2}$ are distinct complex structures on 
$(S,\sigma )$, then $\omega _{J_{1}}$ (resp., $\omega _{J_{2}}$) predicts 
dispersion in $N_{J_{2}}$ (resp., $N_{J_{1}}$).} 

\begin{proof} We shall prove the contrapositive.  Suppose, then, that
  there is some extension $\hat{\omega}_{J_{1}}$ of $\omega _{J_{1}}$ to
  $\mathbf{B}(\hil{F}(\hil{S}_{J_{2}}))$ that is dispersion-free on all bounded
  functions of $N_{J_{2}}$.  Then $\hat{\omega }_{J_{1}}$ is multiplicative
  for the product of the bounded operator $e^{\pm itN_{J_{2}}}$ with any other element of
  $\mathbf{B}(\hil{F}(\hil{S}_{J_{2}}))$ (KR [1997], Ex. 4.6.16).  Hence, 
  by Eqn. (\ref{eq:forty1}),
\begin{eqnarray} \omega _{J_{1}}(W(\cos t +\sin t J_{2}f)) &=& \hat{\omega
      }_{J_{1}} \left( e^{-itN_{J_{2}}}\pi_{\omega _{J_{2}}}(W(f))e^{itN_{J_{2}}} \right) \\
    &=&\hat{\omega}_{J_{1}}(e^{-itN_{J_{2}}})\,\omega _{J_{1}}(W(f))
    \,\hat{\omega}_{J_{1}}(e^{itN_{J_{2}}}) \\
    &=&\omega _{J_{1}}(W(f)), \end{eqnarray} 
for all $f\in S$ and $t\in \mathbb{R}$.  In particular, we may set 
$t=\pi /2$, and it follows that $\omega _{J_{1}}(W(J_{2}f))=\omega _{J_{1}}(W(f))$ 
for all $f\in S$.  Since $e^{-x}$ is a one-to-one function of
    $x\in \mathbb{R}$, it follows from (\ref{eq:vacuum}) that 
\begin{equation}
(f,f)_{J_{1}}=(J_{2}f,J_{2}f)_{J_{1}} , \qquad f\in S , \end{equation}
and $J_{2}$ is a real-linear isometry of the Hilbert space 
$\hil{S}_{J_{1}}$.  We next show that $J_{2}$ is in fact a unitary 
operator on $\hil{S}_{J_{1}}$.

Since $J_{2}$ is a symplectomorphism, 
$\mathrm{Im}(J_{2}f,J_{2}g)_{J_{1}}=\mathrm{Im}(f,g)_{J_{1}}$ 
for any two elements $f,g\in S$.  We 
also have
\begin{equation} \label{eq:blobb}
|f+g|_{J_{1}}^{2} = |f|_{J_{1}}^{2}+|g|_{J_{1}}^{2}+2\mathrm{Re}(f,g)_{J_{1}},
\end{equation}
\begin{eqnarray} 
|J_{2}f+J_{2}g|_{J_{1}}^{2} & = & 
|J_{2}f|_{J_{1}}^{2}+|J_{2}g|_{J_{1}}^{2}+2\mathrm{Re}(J_{2}f,J_{2}g)_{J_{1}} \\ 
\label{eq:blobbo}
& = & 
|f|_{J_{1}}^{2}+|g|_{J_{1}}^{2}+2
\mathrm{Re}(J_{2}f,J_{2}g)_{J_{1}},
\end{eqnarray}
using the fact that $J_{2}$ is isometric.  But $J_{2}(f+g)=J_{2}f+J_{2}g$, since $J_{2}$ is
real-linear.  Thus,
\begin{equation}
 |J_{2}f+J_{2}g|_{J_{1}}^{2} = |J_{2}(f+g)|_{J_{1}}^{2} = |f+g|_{J_{1}}^{2},
 \end{equation}
using again the fact that $J_{2}$ is isometric.  Cancellation with 
Eqns.  (\ref{eq:blobb}) and (\ref{eq:blobbo})
then gives $\mathrm{Re}(f,g)_{J_{1}}=\mathrm{Re}(J_{2}f,J_{2}g)_{J_{1}}$.
Thus, $J_{2}$ preserves the inner product between any two vectors in 
$\hil{S}_{J_{1}}$. All 
 that remains to show is that $J_{2}$ is complex-linear. 
 So let $f\in \hil{S}_{J_{1}}$.  Then, \begin{equation}
(J_{2}(if),J_{2}g)_{J_{1}} = (if,g)_{J_{1}}=-i(f,g)_{J_{1}}=-i(J_{2}f,J_{2}g)_{J_{1}}=(iJ_{2}f,J_{2}g)_{J_{1}} ,\end{equation}
for all $g\in \hil{H}$.  Since $J_{2}$ is onto, it follows that
$(J_{2}(if),g)_{J_{1}}=(iJ_{2}f,g)_{J_{1}}$ for all $g\in \hil{H}$ and therefore $J_{2}(if)=iJ_{2}f$.  

Finally, since $J_{2}$ is unitary and
$J_{2}^{2}=-I$, it follows that $J_{2}=\pm iI=\pm J_{1}$.  However, if $J_{2}=-J_{1}$,
then \begin{equation}
-\sigma (f,J_{1}f)=\sigma (f,J_{2}f)\geq 0 , \qquad f\in S ,\end{equation}
since $J_{2}$ is a complex structure.  Since $J_{1}$ is also a complex
structure, it follows that $\sigma (f,J_{1}f)=0$ for all $f\in S$ and 
$S=\{ 0\}$.  Therefore, $J_{2}=J_{1}$.  \end{proof}

\noindent\textbf{Proposition \ref{fulling}.} \emph{A pair of Fock representations
$\pi_{\omega_{J_{1}}}$, $\pi_{\omega_{J_{2}}}$ are unitarily 
equivalent if and only if
$\omega_{J_{1}}$ assigns $N_{J_{2}}$ a finite value (equivalently, 
$\omega_{J_{2}}$ assigns $N_{J_{1}}$ a finite value).} 
 
 \begin{proof}  $S$ may be thought of as a \emph{real} Hilbert space 
 relative to either of the inner products $\mu_{1},\mu_{2}$ defined by
 \begin{equation} \label{eq:boring}
 \mu_{1,2}(\cdot,\cdot):=\mathrm{Re}(\cdot,\cdot)_{J_{1,2}}
 =\sigma(\cdot,J_{1,2}\cdot). 
 \end{equation}
 We shall use Van Daele and Verbeure's [1971] Theorem 2: 
 $\pi_{\omega_{J_{1}}},\pi_{\omega_{J_{2}}}$ are unitarily equivalent 
 if and only if the positive operator $-[J_{1},J_{2}]_{+}-2I$ on $S$ 
 is a trace-class 
 relative to $\mu_{2}$.  (Since unitary equivalence is symmetric, the 
 same ``if and only if'' holds with $1\leftrightarrow 2$.) 
 
 As we know, we can build any number operator $N_{J_{2}}(f)$ ($f\in S$) on 
 $\hil{H}_{\omega_{J_{1}}}$ by using the complex structure $J_{2}$ in 
 Eqns. (\ref{eq:two}).  In terms of field operators, the result is
 \begin{equation}  \label{eq:ghost}
 N_{J_{2}}(f)=2^{-1}(\Phi(f)^{2}+\Phi(J_{2}f)^{2}+i[\Phi(f),\Phi(J_{2}f)]).
 \end{equation}
 Observe that 
 $N_{J_{2}}(J_{2}f)=N_{J_{2}}(f)$, which had better be the case, since $N_{J_{2}}(f)$ 
 represents the number of $J_{2}$-quanta with wavefunction 
 in the \emph{subspace} of $\hil{S}_{J_{2}}$ generated by $f$.  The 
 expectation value of an arbitrary ``two-point function'' in $J_{1}$-vacuum 
 is given by
 \begin{eqnarray} \label{eq:watchout}
& \langle\Omega_{\omega_{J_{1}}},\phi (f_{1})\phi 
(f_{2})\Omega_{\omega_{J_{1}}}\rangle \\ = &
(-i)^{2}\frac{\partial
^{2}}{\partial t_{1}\partial t_{2}}\omega_{J_{1}}(W(t_{1}f_{1})W(t_{2}f_{2}))
|_{t_{1}=t_{2}=0} \\
=&-\frac{\partial ^{2}}{\partial t_{1}\partial 
t_{2}}\exp(-\frac{1}{2}t_{1}t_{2}(f_{1},f_{2})_{J_{1}}-\frac{1}{4}t_{1}^{2}
(f_{1},f_{1})_{J_{1}}-\frac{1}{4}t_{2}^{2} 
(f_{2},f_{2})_{J_{1}})|_{t_{1}=t_{2}=0} \\ \label{eq:sign}
=&\frac{1}{2}(f_{1},f_{2})_{J_{1}},
\end{eqnarray}
invoking (\ref{eq:reg}) in the first equality, and the Weyl 
relations (\ref{eq:herro}) together with Eqns. (\ref{eq:ip}), 
(\ref{eq:vacuum}) 
to obtain the second.   Plugging Eqn. (\ref{eq:sign}) back into 
(\ref{eq:ghost}) and using (\ref{eq:boring}) eventually yields 
\begin{equation} \label{eq:calculation}
\langle\Omega_{\omega_{J_{1}}},N_{J_{2}}(f)\Omega_{\omega_{J_{1}}}\rangle=   
2^{-2}\mu _{2}(f,(-[J_{1},J_{2}]_{+}-2I)f).
\end{equation}

Next, recall that on the Hilbert space $\hil{H}_{\omega_{J_{2}}}$, $N_{J _{2}}=
\sum_{k=1}^{\infty }N_{J _{2}}(f_{k})$, where $\{f_{k}\}\subseteq 
\hil{S}_{J_{2}}$ is any orthonormal basis.   Let 
$\widehat{\omega}_{J_{1}}$ be any extension of $\omega_{J_{1}}$ to 
$\mathbf{B}(\hil{H}_{\omega_{J_{2}}})$.  The calculation that resulted in 
expression (\ref{eq:calculation}) was done in $\hil{H}_{\omega_{J_{1}}}$, 
however, only finitely many-degrees of freedom were involved.  Thus 
the Stone-von Neumann uniqueness theorem ensures that 
(\ref{eq:calculation}) gives the 
value of each individual $\widehat{\omega}_{J_{1}}(N_{J_{2}}(f_{k}))$. 
Since for any finite $m$, $\sum_{k=1}^{m}N_{J
_{2}}(f_{k})\leq N_{J _{2}}$ as positive operators, we must also have 
\begin{equation}
\sum_{k=1}^{m}\widehat{\omega }_{J_{1}}(N_{J _{2}}(f_{k}))=\widehat{\omega }
_{J_{1}}\left( \sum_{k=1}^{m}N_{J _{2}}(f_{k})\right) \leq \widehat{\omega }%
_{J_{1}}(N_{J_{2}}).
\end{equation}  Thus, $\widehat{\omega }_{J_{1}}(N_{J_{2}})$ will be defined 
just in case the sum 
\begin{equation} \label{eq:sum}
\sum_{k=1}^{\infty }\widehat{
\omega }_{J_{1}}(N_{J _{2}}(f_{k}))=\sum_{k=1}^{\infty}
\widehat{\omega }_{J_{1}}(N_{J
_{2}}(J_{2}f_{k}))
\end{equation}
 converges.  Using (\ref{eq:calculation}), this is, in turn, equivalent to 
\begin{equation} \label{eq:ex}
\sum_{k=1}^{\infty }\mu 
_{2}(f_{k},(-[J_{1},J_{2}]_{+}-2I)f_{k})+\sum_{k=1}^{%
\infty }\mu _{2}(J_{2}f_{k},(-[J_{1},J_{2}]_{+}-2I)J_{2}f_{k})<\infty.
\end{equation}
However, it is easy to see that $\{f_{k}\}$ is a $J_{2}$-orthonormal basis
just in case $\{f_{k},J_{2}f_{k}\}$ forms an orthonormal basis in 
$S$ relative to the inner product $\mu _{2}$. Thus, Eqn. 
(\ref{eq:ex}) is none other than the statement that the operator 
$-[J_{1},J_{2}]_{+}-2I$ on $S$ is trace-class relative to $\mu_{2}$, which is equivalent to 
the unitary equivalence of 
$\pi_{\omega_{J_{1}}},\pi_{\omega_{J_{2}}}$. (The same argument, of 
course, applies with $1\leftrightarrow 2$ throughout.) \end{proof}

\noindent\textbf{Proposition \ref{chaiken}.} \emph{If $\rho$ is a regular state 
of $\alg{W}$ disjoint from 
the Fock state $\omega _{J}$, then $\inf _{F\in \fin} \, \Bigl\{ \,
    \mathrm{Prob}\,^{\rho}(N_{F}\in [0,n]) \Bigr\} =0$ for every $n\in 
    \mathbb{N}$.}

\begin{proof} Suppose that $\omega _{J}$ and $\rho$ are disjoint; i.e.,
  $\mathfrak{F}(\omega _{J})\cap \mathfrak{F}(\rho)=\emptyset$.  
  First, we show that $D(n_{\rho})=\{0\}$, where $n_{\rho}$ is the 
  quadratic form on $\hil{H}_{\rho}$ which, if densely defined, would 
  correspond to the total 
  $J$-quanta number operator.
  
  Suppose, for reductio ad absurdum, that $D(n_{\rho})$ contains
  some unit vector $\psi$.  Let $\omega$ be the state of $\alg{W}$
  defined by
  \begin{equation} \omega (A)=\left\langle \psi ,\pi
      _{\rho}(A)\psi \right\rangle , \qquad A\in \alg{W}.\end{equation} 
Since $\omega \in \mathfrak{F}(\rho)$, it follows that $\omega$ is 
a regular state of $\alg{W}$
    (since $\rho$ itself is regular), and that $\omega \not\in
    \mathfrak{F}(\omega _{J})$.  Let $P$ be the projection onto the 
    closed subspace in $\hil{H}_{\rho}$ generated by the set 
$\pi _{\rho}(\alg{W})\psi$.  If we
  let $P\pi _{\rho}$ denote the subrepresentation of $\pi _{\rho}$ on
    $P\hil{H}_{\rho}$, then $(P\pi _{\rho},P\hil{H}_{\rho})$ is a
    representation of $\alg{W}$ with
    cyclic vector $\psi$.  By the uniqueness of the GNS
  representation, it follows that $(P\pi _{\rho},P\hil{H}_{\rho})$ is 
  unitarily equivalent to
  $(\pi _{\omega},\hil{H}_{\omega})$.  In particular, since 
  $\Omega_{\omega}$ is the image in $\hil{H}_{\omega}$ of 
  $\psi\in P\hil{H}_{\rho}$, $D(n_{\omega})$ contains a vector cyclic 
  for $\pi _{\omega}(\alg{W})$ in $\hil{H}_{\omega}$.  However, by BR ([1996], Thm. 
  4.2.14, $(3)\Rightarrow (1)$), this implies that 
  $\omega \in \mathfrak{F}(\omega _{J})$ --- a contradiction.  Therefore,
  $D(n_{\rho})=\{ 0\}$. 
  
  Now suppose, again for reductio ad absurdum, that 
  \begin{equation}
  \inf _{F\in \fin} \, \Bigl\{ \,
    \mathrm{Prob}\,^{\rho}(N_{F}\in [0,n]) \Bigr\} \not=0. 
    \end{equation} 
    Let $E_{F}:=[E_{\rho}(F)]([0,n])$ and let $E:=\bigwedge
  _{F\in \mathbb{F}} E_{F}$.  
  Since the family $\{ E_{F} \}$ of projections is downward
  directed (i.e., $F\subseteq F'$ implies $E_{F}\geq E_{F'}$), we have
\begin{equation} 0\:\neq \:\inf _{F\in \fin} \{ \langle \Omega
    _{\rho},E_{F}\Omega _{\rho}\rangle \} \: =\: \langle \Omega
    _{\rho},E\Omega _{\rho}\rangle \: =\: \norm{E\Omega _{\rho}}^{2}
    .\end{equation}
Now since $E_{F}E\Omega _{\rho}=E\Omega
  _{\rho}$, it follows that \begin{equation} [n_{\rho}(F)]
  (E\Omega _{\rho})\leq n ,\end{equation} for all $F\in\fin$.  Thus,
    $E\Omega _{\rho}\in D(n_{\rho})$ and $D(n_{\rho})\neq \{ 0\}$ --- 
    contradicting the conclusion of the previous paragraph.  
\end{proof}
  
\section*{References}

\noindent Arageorgis, A. [1995]: \textit{Fields, Particles, and Curvature:
  Foundations and Philosophical Aspects of Quantum Field Theory on
  Curved Spacetime.}  Ph.D. dissertation, University of Pittsburgh.\vspace{.1in}

\noindent Arageorgis, A., Earman, J., and Ruetsche, L. [2000], `Weyling the Time Away:
The Non-Unitary Implementability of Quantum Field Dynamics on Curved
Spacetime and the Algebraic Approach to Quantum Field Theory', 
forthcoming.\vspace{.1in}

\noindent Baez, J., Segal, I., and Zhou, Z. [1992]: {\it Introduction to 
Algebraic and Constructive Quantum Field Theory}, Princeton, NJ: 
Princeton University Press.\vspace{.1in}

\noindent Belinski\u{i}, V. [1997]: `Does the Unruh Effect Exist?', 
{\it Journal of European Theoretical Physics Letters}, {\bf 65}, pp. 
902--908.\vspace{.1in}

\noindent Bratteli, O. and Robinson, D. [1996]: {\it Operator Algebras and
  Quantum Statistical Mechanics}, Vol. 2, NY: Springer.\vspace{.1in}
  
  \noindent Bridgman, P. [1936]: {\it The Nature of Physical Theory}, New 
  York: Dover.\vspace{.1in}
  
  \noindent Busch, P., Grabowski, M., and Lahti, P. [1995]: {\it 
  Operational Quantum Physics}, Berlin: Springer.\vspace{.1in}
  
  \noindent Chaiken, D. [1967]: `Finite-particle representations and states of the
canonical commutation relations', {\it Annals of Physics}, {\bf 42},
pp. 23--80.\vspace{.1in}

\noindent Clifton, R. and Halvorson, H. [2000]: `Entanglement and 
Open Systems in Algebraic Quantum Field Theory', {\it Studies in 
History and Philosophy of Modern Physics}, forthcoming.\vspace{.1in}

\noindent Davies, P. [1984]: `Particles Do Not Exist', in S. Christensen
(ed.), \textit{Quantum Theory of Gravity}, Bristol: 
Adam-Hilger.\vspace{.1in}

\noindent DeWitt, B. [1979a]: `Quantum Field Theory on Curved 
Spacetime', \emph{Physics Reports}, \textbf{19}, pp. 
295--357.\vspace{.1in}

\noindent DeWitt, B. [1979b]: `Quantum Gravity: The New Synthesis', in S.
Hawking et al. (\textit{eds.}), \textit{General Relativity}, Cambridge:
Cambridge University Press, pp. 680--745.\vspace{.1in}

  \noindent Emch, G. [1972]: \textit{Algebraic Methods in Statistical Mechanics and
Quantum Field Theory}, New York: John Wiley.\vspace{.1in}

\noindent Fedotov, A., Mur, V., Narozhny, N., Belinski\u{i}, V., and 
Karnakov, B. [1999]: `Quantum Field Aspect of the Unruh Problem', 
\emph{Physics Letters A}, \textbf{254}, pp. 126--132.\vspace{.1in}

\noindent Fell, J. [1960]: `The Dual Spaces of $C^{*}$-algebras', 
\emph{Transactions of the American Mathematical Society}, 
\textbf{94}, pp. 365--403.\vspace{.1in}

\noindent Fulling, S. [1972]: \textit{Scalar Quantum Field Theory 
in a Closed Universe of Constant Curvature.}  Ph.D. dissertation, 
Princeton University.\vspace{.1in}

\noindent Fulling, S. [1989]: \textit{Aspects of Quantum Field Theory in Curved
Space-time}, Cambridge: Cambridge University Press.\vspace{.1in}

\noindent Gerlach, U. [1989]: `Quantum States of a Field Partitioned by an 
Accelerated Frame', \textit{Physical Review D}, \textbf{40}, pp. 
1037--1047.\vspace{.1in}

\noindent Glymour, C. [1971]: `Theoretical Realism and Theoretical Equivalence', in R. 
Buck and R. Cohen (eds.), \textit{Boston Studies in Philosophy of
Science, VIII}, Dordrecht: Reidel, pp. 275--288.\vspace{.1in}

\noindent Haag, R. and Kastler, D. [1964]: `An Algebraic Approach to Quantum Field
Theory', \textit{Journal of Mathematical Physics}, \textbf{5}, pp. 
848--861.\vspace{.1in}

\noindent Haag, R. [1992]: \textit{Local Quantum Physics}, Berlin: 
Springer-Verlag.\vspace{.1in}

\noindent Halvorson, H. [2000a]: `On the Nature of Continuous Physical Quantities 
in Classical and Quantum
Mechanics', \emph{Journal of Philosophical Logic}, forthcoming.\vspace{.1in}

\noindent Halvorson, H. [2000b]: `Reeh-Schlieder Defeats 
Newton-Wigner: On Alternative Localization Schemes in Relativistic 
Quantum Field Theory', submitted to \textit{Philosophy of 
Science}.\vspace{.1in}

\noindent Halvorson, H. and Clifton, R. [1999]: `Maximal Beable 
Subalgebras of Quantum Mechanical Observables', \emph{International 
Journal of Theoretical Physics}, \textbf{38}, pp. 
2441--2484.\vspace{.1in} 

\noindent Horuzhy, S. [1988] {\it Introduction to Algebraic Quantum Field
  Theory}, Boston: Kluwer. \vspace{.1in} 

\noindent Huggett, N., and Weingard, R. [1996]: `Critical Review: Paul Teller's 
\textit{Interpretive Introduction to Quantum Field Theory'}, \textit{\
Philosophy of Science}, \textbf{63}, pp. 302--314.\vspace{.1in}

\noindent Jauch, J. [1973]: \textit{Are Quanta Real? A Galilean Dialogue},
Bloomington: Indiana University Press.\vspace{.1in}

\noindent Kadison, R. [1965]: `Transformations of States in Operator Theory and 
Dynamics', \textit{Topology}, \textbf{3}, pp. 177--198.\vspace{.1in}

\noindent Kadison, R. and Ringrose, J. [1997]: {\it Fundamentals of the Theory
  of Operator Algebras}, Providence, R.I.: American Mathematical
Society.\vspace{.1in}

\noindent Kay, B. [1979]: `A uniqueness result in the Segal-Weinless approach
to linear Bose fields', {\it Journal of Mathematical Physics}, {\bf
  20}, pp. 1712--1713.\vspace{.1in}

\noindent Kay, B. [1985]: `The double-wedge algebra for quantum fields on
Schwarzschild and Minkowski spacetimes', {\it Communications in
  Mathematical Physics}, {\bf 100}, pp. 57--81.\vspace{.1in}
  
  \noindent Nikoli\'{c}, H. [2000]: `Can the Unruh-DeWitt Detector 
  Extract Energy from the Vacuum?', {\tt hep-th/0005240}.\vspace{.1in}
  
  \noindent Petz, D. [1990]: \textit{An Invitation to the Algebra of 
  the Canonical Commutation Relations}, Belgium: 
  Leuven University Press.\vspace{.1in}
  
 \noindent Redhead, M. [1995]: `The Vacuum in Relativistic Quantum Field Theory',
in \textit{Proceedings of the Philosophy of Science Association 1994}, Vol.
2, pp. 77--87.\vspace{.1in}
  
\noindent Reichenbach, H. [1938]: \textit{Experience and Prediction: An Analysis of
the Foundations and Structure of Knowledge}, Chicago: University of Chicago
Press.\vspace{.1in}

\noindent Roberts, J. and Roepstorff, G. [1969]: `Some Basic Concepts of Algebraic
Quantum Theory', \textit{Communications in Mathematical Physics}
 \textbf{11}, pp. 321-338.\vspace{.1in}

\noindent Robinson, D. [1966]: `Algebraic Aspects of Relativistic Quantum Field
Theory', in M. Chretien and S. Deser (eds.), \textit{Axiomatic Field Theory,
Vol. 1}, New York: Gordon and Breach, pp. 389--516.\vspace{.1in}

\noindent R\"{u}ger, A. [1989]: `Complementarity meets General Relativity: A Study in
Ontological Commitments and Theory Unification', \textit{Synthese}, \textbf{79}, 
pp. 559--580.\vspace{.1in}

\noindent Salmon, M., Earman, J., Glymour, C., Lennox, J., Machamer, 
P., McGuire, J., Norton, J., Salmon, W., and Schaffner, K. [1992]: 
\emph{Introduction to the Philosophy of Science}, New Jersey: 
Prentice-Hall.\vspace{.1in}

\noindent Sciama, D., Candelas, P., and Deutsch, D. [1981]: `Quantum 
Field Theory, Horizons, and Thermodynamics', \textit{Advances in 
Physics}\textbf{30}, 
pp. 327--366.\vspace{.1in}

\noindent Segal, I. [1959]: `Foundations of the Theory of Dynamical Systems of
Infinitely Many Degrees of Freedom, I', {\it Mat. Fys. Medd. Dan. Vid.
  Selsk.}, {\bf 31}, pp. 1--39.\vspace{.1in}
  
  \noindent Segal, I. [1961] `Foundations of the Theory of Dynamical Systems of
Infinitely Many Degrees of Freedom, II', {\it Canadian Journal of 
Mathematics}, {\bf 13}, pp. 
1--18.\vspace{.1in}

\noindent Segal, I. [1963]: \textit{Mathematical Problems of Relativistic Physics},
Providence, RI: American Mathematical Society.\vspace{.1in}

\noindent Segal, I.  [1967]: `Representations of the Canonical Commutation
Relations', in F. Lur\c{c}at (\textit{ed.}), \textit{Carg\`{e}se Lectures in
Theoretical Physics}, New York: Gordon and Breach, pp. 107--170.\vspace{.1in}

\noindent Segal, I. and Goodman, R. [1965]: `Anti-locality of certain
Lorentz-invariant operators', {\it Journal of Mathematics and
  Mechanics}, {\bf 14}, pp. 629--638.\vspace{.1in}

\noindent Summers, S. [1998]: `On the Stone-von Neumann Uniqueness Theorem
and Its Ramifications', to appear in M. R\'{e}dei and M. Stolzner (eds.),
\textit{John von Neumann and the Foundations of Quantum Mechanics.}\vspace{.1in}

\noindent Teller, P. [1995]: \textit{An Interpretive Introduction to Quantum
  Field Theory}, Princeton: Princeton University Press.\vspace{.1in}
  
  \noindent Teller, P. [1996]: `Wave and Particle Concepts in Quantum Field Theory', in
R. Clifton (\textit{ed.}), \textit{Perspectives on Quantum Reality},
Dordrecht: Kluwer Academic Publishers, pp. 143--154.\vspace{.1in}
  
 \noindent Teller, P. [1998]: `On Huggett and Weingard's Review of 
 \textit{An Interpretive Introduction to Quantum Field Theory}: Continuing the
Discussion', \textit{Philosophy of Science}, \textbf{65}, pp.
151--161.\vspace{.1in}

  \noindent Torretti, R. [1999]: \textit{The Philosophy of Physics}, Cambridge: 
  Cambridge University Press.\vspace{.1in}
  
  \noindent Van Daele, A., and Verbeure, A. [1971]: `Unitary 
  Equivalence of Fock Representations on the Weyl Algebra', 
  \emph{Communications in Mathematical Physics}, \textbf{20}, pp. 
  268--278.\vspace{.1in}
  
 \noindent Von Neumann, J. [1931]: `Die Eindeutigkeit der Schr\"{o}dingerschen 
 Operatoren', \emph{Math. Ann.}, {\bf 104}, pp. 570--578.\vspace{.1in}
 
 \noindent Unruh, W. and Wald, R. [1984]: `What Happens When an Accelerating 
 Observer Detects a Rindler Particle?', \emph{Physical Review D}, 
 {\bf 29}, pp. 1047--1056.\vspace{.1in}
  
\noindent Wald, R. [1994]: \textit{Quantum Field Theory in Curved Spacetime and Black
Hole Thermodynamics}, Chicago: University of Chicago Press.\vspace{.1in}

\end{document}